\documentclass[11pt,a4paper,times]{article}
\usepackage[a4paper, margin=1in]{geometry}
\usepackage[polutonikogreek,english]{babel}

\usepackage{graphicx} 
\usepackage{color}
\usepackage{amsmath} 

\usepackage{amssymb}           
\usepackage{amsthm}             
\usepackage{newtxtext}          
\usepackage{newtxmath} 
\usepackage{amsfonts}
\usepackage{comment}
\usepackage{float}
\usepackage{subfig}
\usepackage[mathlines]{lineno}
\usepackage{appendix}

\newcommand{\bO}{\boldsymbol{0}}
\newcommand{\bB}{\boldsymbol{B}}
\newcommand{\bC}{\boldsymbol{C}}
\newcommand{\bE}{\boldsymbol{E}}
\newcommand{\fb}{\boldsymbol{f}}
\newcommand{\bF}{\boldsymbol{F}}
\newcommand{\bI}{\boldsymbol{I}}
\newcommand{\bK}{\boldsymbol{K}}
\newcommand{\bk}{\boldsymbol{k}}
\newcommand{\bH}{\boldsymbol{H}}
\newcommand{\bh}{\boldsymbol{h}}
\newcommand{\bM}{\boldsymbol{M}}
\newcommand{\bm}{\boldsymbol{m}}
\newcommand{\bN}{\boldsymbol{N}}
\newcommand{\bn}{\boldsymbol{n}}
\newcommand{\bP}{\boldsymbol{P}}
\newcommand{\bQ}{\boldsymbol{Q}}
\newcommand{\bq}{\boldsymbol{q}}
\newcommand{\bR}{\boldsymbol{R}}
\newcommand{\bT}{\boldsymbol{T}}
\newcommand{\bs}{\boldsymbol{s}}
\newcommand{\bU}{\boldsymbol{U}}
\newcommand{\bu}{\boldsymbol{u}}
\newcommand{\bv}{\boldsymbol{v}}
\newcommand{\bV}{\boldsymbol{V}}
\newcommand{\bx}{\boldsymbol{x}}
\newcommand{\bX}{\boldsymbol{X}}

\newcommand{\FT}{\bF^{\rm{T}}}
\newcommand{\FInv}{\bF^{-1}}
\newcommand{\FInvT}{\bF^{-\rm{T}}}
\newcommand{\Fdot}{\dot{\bF}}
\newcommand{\CInv}{\boldsymbol{C}^{-1}}

\newcommand{\Kinv}{\bK^{-1}}

\newcommand{\vs}{\bv_{\rm{s}}}
\newcommand{\vf}{\bv_{\rm{f}}}
\newcommand{\va}{\bv_{\alpha}}

\newcommand{\Vf}{\bV_{\rm{f}}}
\newcommand{\Vs}{\bV_{\rm{s}}}

\newcommand{\vvs}{\bu_{\rm{s}}}
\newcommand{\vvf}{\bu_{\rm{f}}}
\newcommand{\vva}{\bu_{\alpha}}
\newcommand{\Vvf}{\bU_{\rm{f}}}
\newcommand{\Vvs}{\bU_{\rm{s}}}
\newcommand{\Vva}{\bU_{\alpha}}

\newcommand{\Ps}{\bP_{\rm{s}}}
\newcommand{\Pf}{\bP_{\rm{f}}}

\newcommand{\Pel}{\bP^{\rm{e}}}

\newcommand{\ms}{\bm_{\rm{s}}}
\newcommand{\mf}{\bm_{\rm{f}}}
\newcommand{\mfd}{\bm_{\rm{fd}}}
\newcommand{\ma}{\bm_{\alpha}}

\newcommand{\A}{\mathscr{A}(t)}
\newcommand{\Bref}{\mathscr{B}}
\newcommand{\Bt}{\mathscr{B}_{\rm{t}}}
\newcommand{\Pe}{\mathscr{P}^{\rm ext}(t)}
\renewcommand{\SS}{{\mathscr{S}}}

\newcommand{\tr}{{\rm{tr\,}}}
\newcommand{\ddiv}{{\rm{div\,}}}
\newcommand{\Div}{{\rm{Div\,}}}
\newcommand{\Grad}{{\rm{Grad\,}}}
\newcommand{\grad}{{\rm{grad\,}}}
\newcommand{\TimeDerivative}{\frac{\rm{d}}{\rm{dt}}}

\newcommand{\phis}{\phi_{\rm{s}}}
\newcommand{\phif}{\phi_{\rm{f}}}
\newcommand{\phia}{\phi_{\alpha}}
\newcommand{\Phis}{\Phi_{\rm{s}}}
\newcommand{\Phif}{\Phi_{\rm{f}}}
\newcommand{\rhos}{\rho_{\rm{s}}}
\newcommand{\rhof}{\rho_{\rm{f}}}
\newcommand{\rhoa}{\rho_{\alpha}}
\newcommand{\varrhos}{\varrho_{\rm{s}}}
\newcommand{\varrhof}{\varrho_{\rm{f}}}
\newcommand{\varrhoa}{\varrho_{\alpha}}
\newcommand{\hphis}{\hat{\phi}_{\rm{s}}}
\newcommand{\hphif}{\hat{\phi}_{\rm{f}}}

\newcommand{\bsigmas}{\boldsymbol{\sigma}_{\rm{s}}}
\newcommand{\bsigmaf}{\boldsymbol{\sigma}_{\rm{f}}}
\newcommand{\bsigmaa}{\boldsymbol{\sigma}_{\alpha}}
\newcommand{\bchi}{\boldsymbol{\chi}}
\newcommand{\bxi}{\boldsymbol{\xi}}
\newcommand{\bXi}{\boldsymbol{\Xi}}

\newcommand{\Palpha}{\bP_{\alpha}}

\newcommand{\bt}{\boldsymbol{t}}

\graphicspath {{\cdot/}{./Figures/}}

\begin{document}


\title{A model of the human cornea as a hydrated, fluid-saturated medium}

\author{Alessandro Giammarini$^{1}$, Anna Pandolfi $^{1}$\footnote{Corresponding author: Anna Pandolfi, Dipartimento di Ingegneria Civile e Ambientale, Politecnico di Milano - 
Piazza Leonardo da Vinci 32, Milano (MI), 20133, Italy.\\ Email: anna.pandolfi@polimi.it} }

\maketitle

\noindent

{}$^{1}${Dipartimento di Ingegneria Civile e Ambientale, Politecnico di Milano - Piazza Leonardo da Vinci 32, Milano (MI), 20133, Italy.}

\begin{abstract}
We introduce a new model of the human corneal stroma, regarded as a fluid-saturated continuum, able to describe surface flattening and thickness thinning observed in several pathological conditions. In contrast with more common approaches that describe the human cornea as a quasi-incompressible hyperelastic medium, eventually including micro-structured anisotropy and heterogeneity, here we focus on the multi-phase nature of the tissue, where the content of water reaches about 78\% in weight. The study is motivated by the fact that, although purely mechanical continuum models have been demonstrated to be satisfactory and accurate at predicting physiological behaviors, they have not been able to capture the geometrical features of tissue degeneration clearly associated to a reduction of the fluid content in the stroma, such thinning and loss of curvature.
Here, we model the cornea as a fully saturated mixture of a solid phase and a fluid phase, in principle without restricting the formulation to specific assumptions on the actual inhomogeneous nature of both phases. The focus of the study is to understand whether a multiphysics model is capable of explaining, in terms of fluid flux imbalance, such as ectasia and keratoconus. As a first attempt in this direction, we make simple isotropic constitutive assumptions for both phases.
\end{abstract}



\textbf{Key words}: porous materials, solid and fluid mixture,
human cornea, keratoconus, corneal ectasia

\section{Introduction}

The cornea is the transparent, avascular external lens of the eye that provides most of its refractive power. It consists of five distinct layers:  cellular monolayer epithelium, Bowman’s layer, collageneous stroma, Descemet’s membrane, and cellular multilayer endothelium. The corneal stroma is the thickest layer, with structural functions, and is a hydrated, fibrous connective tissue composed primarily of regularly arranged collagen lamellae embedded in a proteoglycan-rich extracellular matrix, interspersed with keratocytes and permeated by interstitial fluid. Traditionally, the cornea has been modeled using continuum mechanics, starting from early contributions \cite{bryant:1996a} based on linear isotropic elasticity, and advancing up to produce realistic models accounting for anisotropy, heterogeneity and nonlinearities, see, e.~g., \cite{whitford:2015, arizagracia:2016, montanino:2018}. 
The most predictive continuum models embed in a phenomenological way the microstructural characteristics of the cornea, such as the organized collagen lamellae in the stroma. These models, often using a patient-specific geometry, have been used to capture the mechanical response of the eye under various conditions, including static and dynamic tests \cite{arizagracia:2015, simonini:2016}, refractive surgery \cite{sanchez:2014}, therapeutic crosslinking \cite{nambiar:2022, bonfanti:2023}.

As far as physiological conditions are concerned, traditional continuum models have provided a sound foundation for understanding corneal mechanics. The physiological state implies the attainment of stable equilibrium for the processes taking place in the eye, including all-important transport phenomena; thus, a purely mechanical approach may be adopted to describe the physiological biomechanical response of the eye, disregarding the presence of other complex physics. In this respect, also the evaluation of the long term modification of the tissue engagement induced by refractive surgery and refractive implants, i.~e., when the homeostatic state has been newly reached, can be considered sufficiently precise. Regrettably, not all eye conditions can be modeled using purely mechanical models and standard stress analysis approaches. Some pathologies are characterized by the degeneration of the complex interplay between cornea solid matrix and interstitial fluid, suggesting poroelasticity as the promising direction for capturing the modified interactions. 

The current literature on eye biomechanics presents a small number of contributions that regard the cornea as a porous medium. One of the first studies describing the cornea as a biphasic medium aimed at improving the understanding of refractive surgery outcomes, and modeled the stroma as a two-phase mixture of solid and fluid, linking the biomechanical stresses with changes in stromal hydration \cite{bryant:1996b}. 
A biphasic model of stromal tissue was used to investigate the compressive biomechanical properties of the cornea \cite{hatami:2013}, seen as a soft, hydrated solid-fluid mixture in line with existing models developed for cartilage \cite{mow:1980, holmes:1990}. A pioneering micro-structural model of the in vivo cornea, described as a solid-fluid composite with interaction between ions and collagen, was proposed with the aim of describing stromal swelling in \cite{cheng:2015}. 
More recently, porous material models have been used to capture the mechanical behavior of bioengineered artificial corneas, that mimic the native environment replicating the biological structure of the cornea, for applications in tissue engineering and regenerative medicine \cite{mahmoudsalehi:2025}. 

Despite these attempts, to date there is no a mature, fully-validated poromechanical continuum model of the human cornea that combines fluid flow and solid deformation within a finite element (FE) framework. The existing studies are predominantly theoretical (based on homogenized formulations) and devoted to interpret the experimental evidence based on inflation and swelling data. 

In the field of human eyes, poromechanical theories may play an important role in describing conditions that fall outside physiological conditions. For one, the evolution of keratoconus represents a case of interest in which a biphasic theory can be of aid in capturing features that cannot be reproduced using more traditional monophasic models. 
Keratoconus is a degenerative pathology that affects progressively the stiffness of the stroma, in later stages characterized by a conic bulging that compromises the refractive and biological functionality of the cornea. The insurgence of keratoconus is associated with a thinning phenomenon which may not be limited to the cone region. Previous studies successfully put in correlation the bulging of the cone with a reduction of the cornea's elastic properties, but were unable to capture the thinning of the cornea to the extent observed experimentally, which can reach 30-50\% of the cornea's thickness \cite{pandolfi:2024, miller:2025}. Another important application of biphasic theories in the human cornea is the modeling of corneal swelling, a phenomenon that can be observed in association to a failure of the endothelium function. With a balanced pump-leak process, the endothelium regulates the flux of fluids and ions from the aqueous humor to the stroma \cite{vanone:2025}. Indeed, biphasic theories are not new in theoretical and numerical applications concerning hydrated biological tissues, see, e.g., growth and remodeling in cartilage \cite{ateshian:2007,Ambrosi2010a,grillo:2012a,pierce:2013,ateshian:2017,Crevacore2019a}. 

In this work, we use a biphasic (solid and fluid) model founded on the mixture theory  \cite{mow:1980, holmes:1990,ateshian:2007,grillo:2012a,Byrne2003a,giammarini:2025} to model the poromechanical behavior of the human stroma in pathologic conditions. Our goal is to capture localized thinning and asymmetric bulging and flattening of the cornea observed in advanced keratoconus, features difficult to achieve using standard purely mechanical models, typically characterized by incompressibility \cite{pandolfi:2024, miller:2025}. As a first step in the development of a realistic evolution model, here we disregard the complex micro-mechanical organization of the stroma, its anisotropy and heterogeneity, and assume an simplified yet instructive hyperelastic isotropic behavior for the solid fraction and homogeneity for the fluid fraction. 

The organization of the paper is as follows. In Section~\ref{sec:GoverningEquations} we describe the governing equations and the parameters of the mixture model. In Section~\ref{sec:FiniteElements} we formulate the balance equations in weak form and introduce the general finite element discretization. In Section~\ref{sec:numericalmodelling} we describe the features of the mesh employed in the simulations, and the modeling strategies concerning the boundary conditions and the characterization of the keratoconus. In Section~\ref{sec:Results} we apply the biphasic model to the simulation of the outcomes of keratoconus starting from a healthy patient-specific geometry of the cornea. In Section~\ref{sec:Conclusions} we comment critically the results of the simulations, discussing benefits and limits of the model with respect to more traditional approaches.

\section{The governing equations of a biphasic medium}
\label{sec:GoverningEquations}

We begin by discussing mass balance and linear momentum equations that govern the mechanics of an isotropic hydrated and fluid-saturated biological porous medium.
We adopt the standard framework of mixture theories \cite{ateshian:2007,grillo:2012a,Byrne2003a}, and do not distinguish the regions occupied by the solid and fluid phases within the tissue. The macroscopic behavior is modelled as the one of a composite material, where each material point is characterized by a local volumetric fraction of solid components and a corresponding volumetric fraction of interstitial fluids. We regard the solid part of the tissue (consisting in fibers, extracellular matrix, and cells) and fluid part (consisting in water and ions) as homogenized substances. 

We consider a reference (or material) configuration $\Bref$ with material coordinates $\bX$ and denote  the regular motion of the solid phase with $\bchi(\bX,t) :\Bref \mapsto \mathbb{R}^{3}$.
The current (or spatial) configuration at time $t$ with spatial coordinates $\bx$ is thus $\Bt = \bchi(\Bref,t)$. The deformation gradient $\bF$ associated with the motion of the solid phase and its determinant $J$ are given by
\begin{equation}
    \bF(\bX,t)=\frac{\partial \bchi (\bX,t)}{\partial \bX}
    \, ,
    \qquad
    J = \det{\bF(\bX,t)}
    \,.
    \label{eq:deformationGradient}
\end{equation}
To describe the mixture components, we introduce the intrinsic (true) densities $\varrhos(\bx,t) $ and $\varrhof(\bx,t) $ of solid and fluid phases, respectively, and the volume ratios $\phis(\bx,t) $ e $\phif(\bx,t)$ as the percentage of volume occupied by solid and fluid phases within the spatial volumetric element of the mixture. Thus, we define the apparent densities $ \rhos(\bx,t)$ and $ \rhof(\bx,t)$ of the two phases as the product of true densities and volumetric fractions
\begin{equation}
    \rhos(\bx,t) = \phis(\bx,t) \, \varrhos(\bx,t) 
    \, ,
    \qquad
    \rhof(\bx,t) = \phif(\bx,t) \varrhof(\bx,t) 
    \,,
    \label{eq:apparentDensities}
\end{equation}
and the density of the mixture follows as
\begin{equation}
    \rho = \rhos + \rhof =
    \phis \varrhos  +
    \phif \varrhof  \, .
    \label{eq:mixtureDensity}
\end{equation}
The fields $\vs(\bx,t)$ and $\vf(\bx,t)$ denote the spatial velocity of solid and fluid phases, respectively, and $\Vs(\bX,t) = (\vs \circ \bchi)(\bX,t)$ and $\Vf(\bX,t) = (\vf \circ \bchi)(\bX,t)$ denote the composition of the velocities with the motion of the solid phase. The velocity of the mixture is defined as 
\begin{equation}
    \bv = \phif \vf +\phis \vs \, . 
    \label{eq:mixtureVelocity}
\end{equation}

\begin{figure}[H]
    \centering
    \includegraphics[width=0.75\linewidth]{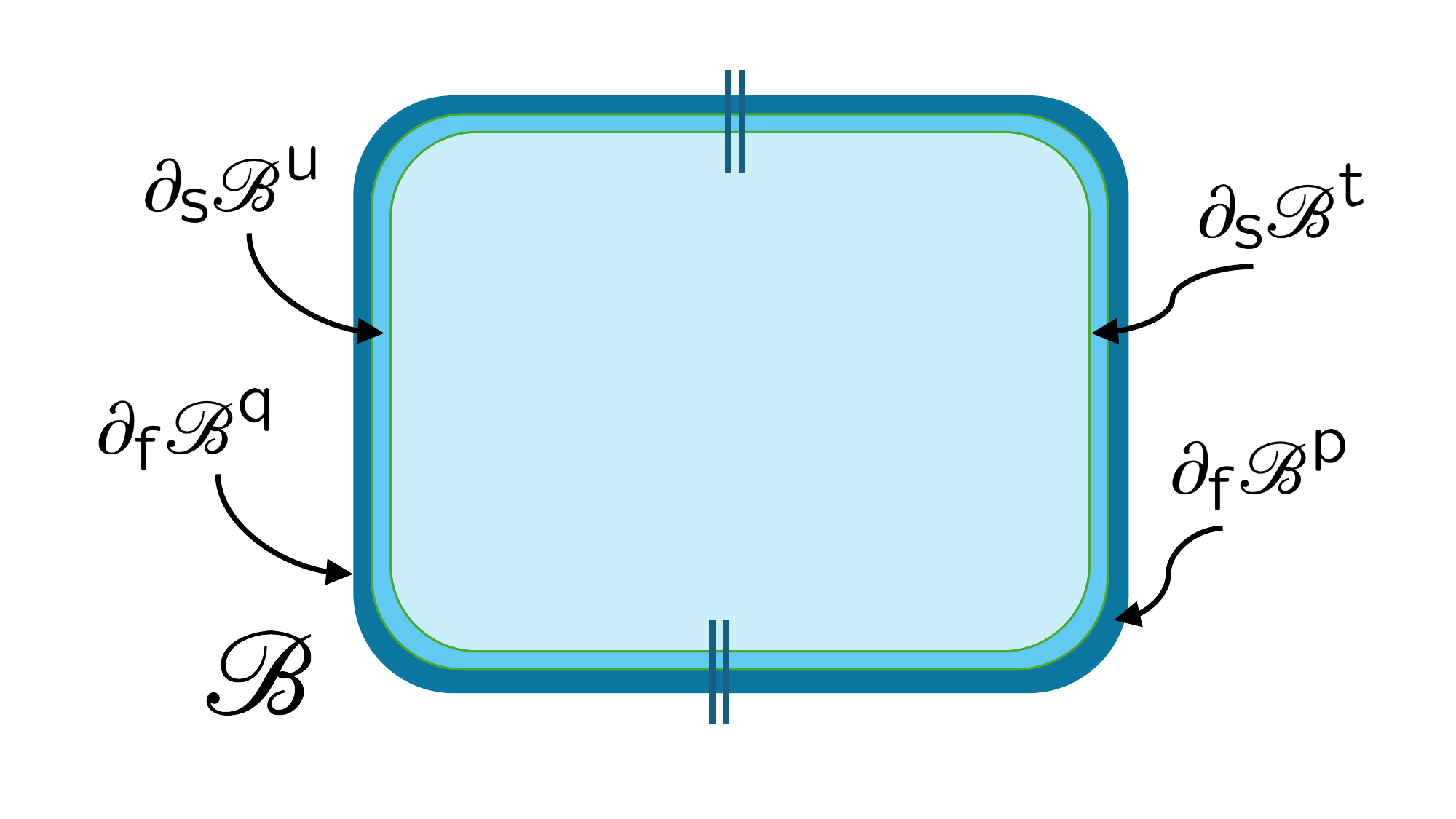}
    \caption{\footnotesize A generic biphasic body, occupying the domain $\Bref$, with boundaries of different types. Light blue border denotes the solid boundaries: $\partial_{\rm s} \Bref^{\rm u}$ is the Dirichlet type boundary, with imposed displacements, and $\partial_{\rm s} \Bref^{\rm t}$ is the Neumann type boundary, with imposed tractions. Dark blue border denotes the fluid boundaries: $\partial_{\rm f} \Bref^{\rm p}$ is the Dirichlet type boundary, with imposed pressures, and $\partial_{\rm f} \Bref^{\rm q}$ is the Neumann type boundary, with imposed fluxes.}  
    \label{fig:fig1}
\end{figure}

For each phase, the boundary of the system $\partial \Bref$ is disjoined in two distinct portions \cite{Bonet2008,Giammarini2025b}, one of Dirichlet type, where the primary variables are imposed, and the other of Neumann type, where the gradient of the primary variables are imposed.
Therefore, we distinguish between $\partial_{\rm s} \Bref^{\rm u}$ and $\partial_{\rm s} \Bref^{\rm t}$ for the solid phase and $\partial_{\rm f} \Bref^{\rm p}$ and $\partial_{\rm f} \Bref^{\rm q}$ for the fluid phase, see Fig.~\ref{fig:fig1}.

\subsection{Mass balance}
\label{sse:massBalance}

In deriving the mass balance equations we will make a few constitutive ansatzs that are suitable for the corneal tissue.  We exclude growth and resorption, assume intrinsic incompressibility and constant true density for both phases, and assume full saturation of the medium \cite{ateshian:2007}.

The spatial form of the global mass balance equations for fluid and solid phases can be expressed as
\begin{equation}
    \frac{d}{dt}\int_{\mathscr{V}_t} \rhoa \, dv = \int_{\mathscr{V}_t} r_{\alpha} \, dv + \int_{\partial \mathscr{V}_t} \rhoa (\bv - \va) \cdot \bn da \qquad (\alpha = \rm{f}, \rm{s}) \, ,
    \label{MassBalance_1}
\end{equation}
where $\mathscr{V}_t \subset \Bt$ is an arbitrary volume in $\Bt$, $r_{\alpha}$ describes the source/sink term, $\bv$ denotes the velocity of the mixture, and $\rhoa (\bv - \va)$ is a compensatory flux relative to the mixture velocity accounting for the leakage of fluid and solid phases.
\\
Excluding growth and absorption ($r_{\alpha} = 0$), the mass balance in terms of apparent densities can be reformulated in local form as
\begin{equation}
    \partial_{t} (\varrhoa\phia)  + \ddiv(\varrhoa\phia \va) = 0 
     \qquad (\alpha = \rm{f}, \rm{s}) \,,
    \label{MassBalance_2}
\end{equation}
where $\rm{div}$ denotes the spatial divergence operator. Exploiting the material derivative with respect to the velocity of the $\alpha$-th phase, see Eq.~\eqref{eq:Appendix1} in Appendix A, we write
\begin{equation}
    \phia\rm{D}_{\alpha}\varrhoa + \varrhoa \big[\partial_{t} \phia  + \ddiv(\phia \va)\big] = 0 \qquad (\alpha = \rm{f}, \rm{s}) \,.
    \label{MassBalance_3}
\end{equation}
Intrinsic incompressibility requires $\rm{D}_{\alpha} \varrhoa = 0$, thus the term $\varrhoa$ drops from Eqs.~\eqref{MassBalance_3}, leading to
\begin{equation}
    \partial_{t} \phia  + \ddiv(\phia \va) = 0 
    \qquad (\alpha = \rm{f}, \rm{s}) \,.
    \label{MassBalance_4}
\end{equation}
The mass balance equation for the mixture is obtained by the sum of Eqs.~\eqref{MassBalance_4}:
\begin{equation}
    \partial_{t} (\phif+\phis)  + \ddiv\big(\phif \vf + \phis\vs\big) = 0 \,.
    \label{MassBalance_5}
\end{equation}
Under saturation condition, i.~e., when the fluid phase fills all pores of the solid phase and
$$\phis + \phif = 1 \,,$$ 
the mass balance states the incompressibility condition for mixtures
\begin{equation}
    \ddiv\big(\phif \vf + \phis\vs\big) = 0 \, .
    \label{MassBalance_6}
\end{equation}
Using the \textit{filtration velocity}, which weights the velocity of the fluid phase relative to the solid phase as
$$\bq= \phif (\vf-\vs)\,,$$ 
Eq.~\eqref{MassBalance_6} can be written alternatively in the form 
\begin{align}
    &\ddiv\vs + \ddiv \bq= 0 \,.
    \label{MassBalance_7}
\end{align}
Eq.~\eqref{MassBalance_7} recasts in a clear form the saturation condition and the mass balance of the fluid–solid mixture. The first term accounts for volumetric changes in the solid phase, while the second considers the relative motion of the fluid with respect to the solid matrix. When the fluid volume fraction is zero (or negligibly small), the equation reduces to the classic incompressibility condition for solids, $\ddiv \vs = 0$. The key idea enforced in Eq.~\eqref{MassBalance_7} is that in a saturated mixture the net volume change is zero, thus a compression of the solid ($\ddiv \vs < 0$) must be exactly balanced by an outflow of fluid ($\ddiv \bq> 0$).

The local form of the mass balance equation in the reference configuration is obtained through standard pull-back operations, as 
\begin{subequations}
    \begin{align}
    & \TimeDerivative\Phis = 0,
    \label{MassBalance_8a}\\
    & \dot{J} + \Div \bQ = 0,
    \label{MassBalance_8b}
    \end{align}
\end{subequations}
where $\Phi_{\rm{s}}= J \phis$ is the referential percentage volume of the solid phase and $\bQ = J\FInv \bq$ is the referential filtration velocity. The incompressibility condition in referential form reads 
\begin{equation}
    \Div[(\Phi_{\rm{s}}\Vs+\Phi_{\rm{f}}\Vf)\, \FInvT] = 0 \, .
    \label{IncompressibilityBis}
\end{equation}
 
\subsection{Linear momentum balance}
\label{ssec:linearMomentum}

We state the linear momentum balance in quasi-static conditions through the virtual power identity. In the current configuration, the system is subject to body forces per unit of mass and surface tractions on the Neumann boundary $\partial\Bt$. We introduce two virtual velocity fields $\vvs$ and $\vvf$, and for each phase $\alpha = \rm{f}, \rm{s}$ we write the virtual power identity as 
\begin{equation}
    \int_{\Bt} \bigg\{ \ma \cdot \vva  + \bsigmaa : \grad \vva \bigg\} \, dv 
    = \int_{\Bt} \fb_{\alpha} \cdot \vva \, dv + \int_{\partial_{\alpha}\Bt}  \bt_{\alpha} \cdot \vva \, da \, ,
    \label{VirtualPower1}
\end{equation}
where $\ma$ denotes the body force representative of solid-fluid interactions,
$\boldsymbol{\sigma}_{\alpha}$ the Cauchy stress tensor of the phase $\alpha$, $\fb_{\alpha}$ the body forces on the phase $\alpha$, 
and $\bt_{\alpha}$ the external tractions on the phase $\alpha$ acting on the Neumann boundary associated with the phase $\partial_{\alpha}\Bt$. Symbols $(\cdot)$ and $(:)$ denote the scalar and the contraction product, respectively. Using Cauchy's theorem
$$\bt_{\alpha} = \boldsymbol{\sigma}_{\!\alpha} \bn \qquad (\alpha=\rm{f},\rm{s})\,,$$
and applying integration by parts and divergence theorem, under the assumption of negligible body forces Eq.~\eqref{VirtualPower1} reduces to the local form 
\begin{equation}
    \ddiv \boldsymbol{\sigma}_{\alpha} - \ma = \bO \qquad \rm{in} \quad \Bt \qquad (\alpha=\rm{f},\rm{s}) \, .
    \label{MomentumBalance1}
\end{equation}

The linear momentum balance equations include the contribution of the Cauchy stress relative to each phase $\boldsymbol{\sigma}_{\alpha}$ and the exchange of linear momentum between the two phases $\ma$.
It is customary to assume that the mixture is closed with respect to exchange of linear momentum between the two phases, thus
\begin{equation}
    \ms + \mf = \bO
    \label{actionReaction}
\end{equation} 
and summing the two linear momentum balance equation we obtain 
\begin{equation}
     \ddiv \left( \boldsymbol{\sigma}_{\rm{s}} + \boldsymbol{\sigma}_{\rm{f}} \right)  =\bO \, \qquad \rm{in} \quad \Bt.
    \label{MomentumBalance2}
\end{equation}
In the reference configuration the virtual fields are denoted with $\Vvs$ and $\Vvf$ and the virtual power identity reads
\begin{equation}
    \int_{\Bref} \bigg\{ J \, \ma \cdot \Vva + \bP_{\alpha}:\Grad \Vva \bigg\} \, dv \, = \int_{\Bref} \boldsymbol{F}_{\alpha} \cdot \Vva \, dV + \int_{\partial_{\alpha}\Bref} \bT_{\alpha} \cdot \Vva \, dA \, ,
    \label{VirtualPower2}
\end{equation}
where $\bP_{\alpha} = J \boldsymbol{\sigma}_{\alpha} \FInvT$ denotes the first Piola-Kirchhoff stress tensor of the phase $\alpha$, $\boldsymbol{F}_{\alpha} = J \fb_{\alpha}$ the referential body forces, and $\bT_{\alpha} dA = \bt_{\alpha} da$ the referential external tractions. The linear momentum balance equations in the reference configuration are  
\begin{equation}
    \Div \Palpha - J \ma = \bO \qquad \rm{in} \quad \Bref  
    \qquad (\alpha=\rm{f},\rm{s}) \, .    \label{MomentumBalance3} 
\end{equation}
Summing the two equations \eqref{MomentumBalance3} and using Eq.~\eqref{actionReaction} we obtain
\begin{equation}
    \Div (\Pf+\Ps) = \bO \qquad \rm{in} \quad \Bref \, .
    \label{MomentumBalance4}
\end{equation}

\subsection{Dissipation inequality}
\label{ssec:dissipationInequality}

We take a thermodynamically consistent approach and complete the set of the governing equations with the dissipation inequality \cite{grillo:2012a,hassanizadeh:1986b,bennethum:2000a}. Under isothermal conditions the internal energy matches the strain energy and the only sources of dissipation of the bi-phasic system under consideration are the mechanical interactions between phases. In the following derivations, we assume that the strain energy density of the fluid phase remains constant, and the solid phase is hyperelastic and homogeneous. The adoption of more complex material behaviors follows the same conceptual derivation.

The dissipation rate $\mathcal{D}(t)$ over an arbitrary volume $\mathscr{V}_{t} \subset \Bt$ is
\begin{equation}
    \mathcal{D}(t) = - \frac{\rm{d}}{\rm{d}t}\A + \Pe + \int_{\partial \mathscr{V}_{\rm  t}} \bh_{\mathscr{A}} \cdot \bn da, \quad {\rm for} \,\, t \ge 0 \, ,\label{DissipationInequality1}
\end{equation}
where $\A$ is the total Helmholtz free energy, $\Pe$ is the external power, and $\bh_{\mathscr{A}}$ is the compensatory flux of energy carried by the velocity mismatch between the two phases, see  \cite{hassanizadeh:1986b, bennethum:2000a, grillo:2012a}. By excluding the presence of mass sources, the dissipation inequality reads 
\begin{align}
    \mathcal{D}(t) = & - \frac{\rm{d}}{\rm{d}t}\bigg\{ \int_{\mathscr{V}_{t}} \left( \phis \psi_{\rm{s}} + \phif \psi_{\rm{f}} \right) \, dv \bigg \} + \int_{\partial \mathscr{V}_{t}} (\bsigmas \vs + \bsigmaf \vf) \cdot \bn da \nonumber \\
    & \quad +\int_{\partial\mathscr{V}_{t}} \big(\phis \, \psi_{\rm{s}} (\bv - \vs) + \phif \, \psi_{\rm{f}} (\bv - \vf) \big) \cdot  \bn da \ge 0,
\label{DissipationInequality2}
\end{align}
where $\psi_{\rm \alpha}$ the Helmholtz free energy density of the phase $\alpha$, with $\alpha = \rm{s},\rm{f}$ (for the solid phase it coincides with the strain energy density) and $\bv$ denotes the velocity of the mixture. An application of the divergence theorem to the second term of the right-hand-side of Eq.~\eqref{DissipationInequality2} combined with the linear momentum balance, Eq.~\eqref{MomentumBalance1}, gives
\begin{align}
    \mathcal{D}(t) = & -\frac{\rm{d}}{\rm{d}t}\bigg\{ \int_{\mathscr{V}_{t}} \left( \phis \psi_{\rm{s}} + \phif \psi_{\rm{f}} \right) dv \bigg \}  + \int_{\partial\mathscr{V}_{t}} \big(\phis \, \psi_{\rm{s}} (\bv - \vs) + \phif \, \psi_{\rm{f}} (\bv - \vf) \big) \cdot  \bn da \nonumber \\
    & + \int_{\mathscr{V}_{t}} \left( \bsigmas :\grad\vs + \ms \cdot \vs + \bsigmaf : \grad\vf + \mf \cdot \vf \right) dv \ge 0 \, .
\label{DissipationInequality3}
\end{align}
Through the Reynold's theorem Eq.~\eqref{eq:Appendix2} and Eq.~\eqref{actionReaction}, the dissipation inequality becomes 
\begin{align}
    \mathcal{D}(t) = 
    & - \int_{\mathscr{V}_{t}} \big[ \partial_{\rm{t}}\big(\phis \psi_{\rm{s}} + \phif \psi_{\rm{f}} \big) + \ddiv \big( \phis \psi_{\rm{s}} \vs + \phi_{\rm{f}}\psi_{\rm{f}} \vf \big) \, \nonumber \\
    & - \bsigmas : \grad\vs + \bsigmaf : \grad\vf + \mf \cdot \left( \vf - \vs \right) \big] dv  \ge 0.
\label{DissipationInequality4}
\end{align}
The local form of the dissipation inequality reads 
\begin{align}
    \Gamma (t) = & - \partial_{\rm{t}} \big(\phis \psi_{\rm{s}} + \phif  \psi_{\rm{f}} \big) - \ddiv \big( \phis \psi_{\rm{s}} \vs + \phi_{\rm{f}}\psi_{\rm{f}} \vf \big)  \nonumber \\  
    &  + \bsigmas : \grad\vs + \bsigmaf : \grad\vf + \mf \cdot \left( \vf - \vs \right) \ge 0 \, .
\label{DissipationInequality5}
\end{align}
If we assume that the Helmholtz energy density of the fluid phase $\psi_{\rm{f}}$ remains constant during the process, we have
\begin{equation}
    \partial_{\rm{t}}(\phif \psi_{\rm{f}} ) + \ddiv (\phi_{\rm{f}}\psi_{\rm{f}} \vf ) = \psi_{\rm{f}} \left( \partial_{\rm{t}}\phif + \ddiv (\phi_{\rm{f}} \vf ) \right) = 0 \, , 
    \label{isothermalConditions}
\end{equation}
which is null because of the fluid phase mass balance, Eq.~\eqref{MassBalance_4}, and drops out of Eq.~\eqref{DissipationInequality5}, that, using the material derivative Eq.~\eqref{eq:Appendix1}, becomes
\begin{align}
    \Gamma(t) = &
    - D_{\rm{s}} (\phis \psi_{\rm{s}} ) - \phis \psi_{\rm{s}} \ddiv \vs \nonumber \\ &
     + \bsigmas : \grad\vs + \bsigmaf : \grad\vf + \mf \cdot \left( \vf - \vs \right) \ge 0 \, .
    \label{DissipationInequality6}    
\end{align}
The expression of the local rate of dissipation in terms of referential quantities becomes
\begin{equation}
    \Gamma (t) = 
    -\Phis
    \frac{\rm{d} \, \psi_{\rm{s}} }{\rm{d}t} 
    + \Ps:\Grad\Vs + \Pf:\Grad\Vf +  J \mf \cdot \left( \Vf -\Vs \right) \ge 0 \, .
\label{DissipationInequality7}
\end{equation}
Due to the elastic nature of the solid phase, the strain energy density rate gives the equilibrium stress $\Pel$ as
\begin{equation}
    \frac{\rm{d} \, \psi_{\rm{s}}(\bF)}{\rm{d}t} 
    = \frac{\partial \, \psi_{\rm{s}}(\bF) }{\partial\bF} : \dot{\bF} = \Pel : \dot{\bF}\, , 
    \label{DerivativePsi}
\end{equation}
and, for the identity $\dot{\bF} = \Grad \Vs$ we obtain
\begin{align}
    \Gamma (t) =  \left(\Ps -\Phis \Pel \right) : \dot{\bF}  + \Pf:\Grad\Vf + J \mf \cdot \left( \Vf -\Vs \right) \ge 0 \, .
\label{DissipationInequality8}
\end{align}
The incompressibility condition defined in Eq.~\eqref{IncompressibilityBis} is finally enforced by augmenting the local dissipation rate with the energetic contribution due to the internal pressure, represented by a Lagrangian multiplier $p$ \cite{liu:1972}. The conditions introduces three additional contributions (see Appendix C for the derivation):
\begin{align}
    \Gamma (t) = &  
    \left\{ \Ps - \Phis \Pel  + p\Phis \FInvT \right\} : \Fdot + \left\{\bP_{\rm{f}} + p\Phif \FInvT  \right\}:\grad\Vf \nonumber \\ 
    &+ \left\{ J\mf + p J \Grad \frac{\Phis}{J}  \FInv \right \} \left( \Vf -\Vs \right) \ge 0
    \label{DissipationInequality9}
\end{align}
From the dissipation inequality we obtain the constitutive relations 
\begin{subequations}
\label{Constitutive}
\begin{align}
    & \Ps = \left(\Pel - p \FInvT \right) \Phis \, ,
    \label{FirstPiolaKirchhoffSolid}\\
    & \Pf = - p  \FInvT\Phif \, ,
    \label{FirstPiolaKirchhoffFluid}\\
    & \mf = -p \Grad \frac{\Phis}{J} \FInv  +
    \phif \FInvT \Kinv \bQ \, .
\label{ExchangeImpulseFluid}
\end{align}
\end{subequations}
Eqs.~\eqref{FirstPiolaKirchhoffSolid}-\eqref{FirstPiolaKirchhoffFluid} follow directly from the assumptions of elastic solid and inviscid fluid. Eq.~\eqref{ExchangeImpulseFluid} is the constitutive law for $\mf$ which accounts for the dissipative nature of the solid-fluid interactions, as the simplest characterization satisfying the residual dissipation inequality (see Appendix C).

Using the components of the stress in Eq.~\eqref{Constitutive}, the linear momentum balance in referential local form \eqref{MomentumBalance4} becomes
\begin{equation}
    \Div (\Phis\Pel - J p \FInvT) = \bO \qquad \rm{in} \, \Bref \, .
    \label{MomentumBalance5}
\end{equation}
The referential local form of the governing equation becomes
\begin{subequations}
\begin{align}
    & \dot{J} + \Div \bQ = 0 && \rm{in} \, \Bref,
    \label{EquationContinuity} \\
    & \Div(\Phis \Pel - J p \FInvT) = \bO && \rm{in} \, \Bref \, .
    \label{EquationEquilibrium} 
\end{align}
\end{subequations}
The problem must be completed with the boundary conditions, which in general form can be stated as
\begin{subequations}
\begin{align}
    & \bU = \bar{\bU} && \rm{on} \quad  \partial_{\rm s} \Bref^{\rm u} \, ,
    \label{BoundaryDirichletSolid} \\
    & \Ps \bN = \bar{\bT} && \rm{on} \quad \partial_{\rm s} \Bref^{\rm t} \, ,
    \label{BoundaryNeumannSolid} \\
    & p = \bar{p} && \rm{on} \quad \partial_{\rm f}  \Bref^{\rm p}\, ,
    \label{BoundaryDirichletFluid} \\
    & \bQ \cdot \bN = \bar{Q} && \rm{on} \quad \partial_{\rm f}  \Bref^{\rm q} \, ,
    \label{BoundaryNeumannFluid} 
\end{align}
\label{Boundaries} 
\end{subequations}
where the barred quantities denote assigned values of the variables (see e.g. \cite{Giammarini2025b}).

\subsection{Constitutive models}
\label{ssec:constitutive}

In the view of numerical simulations, we characterize the material for solid and fluid phases with very simple models.
The solid phase is described by a weakly compressible Mooney--Rivlin elastic model, with  standard decoupling of volumetric and distortional contributions \cite{Bonet2008}
\begin{equation}
    \Psi_{\rm{s}}(\bX) =  \frac{k}{2} (\log J)^{2} +  \frac{1}{2}\mu_{1} (\bar{I}_{1}-3) + \frac{1}{2}\mu_{2} (\bar{I}_{2}-3)  
    \label{MRElasticEnergy}
\end{equation}
where $\mu = \mu_1 + \mu_2$ is the shear modulus of the material and $\bar{I}_1$ and $\bar{I}_2$ are the first and the second invariant, respectively, of the distortional right Cauchy--Green tensor $\bar{\bC} = J^{-2/3}\FT \bF$
\begin{subequations}
\begin{align}
    & \bar{I}_{1} = J^{-2/3} \tr \bC
    \label{FirstInvariant}\\    
    & \bar{I}_{2} = J^{-4/3} \frac{1}{2}\big( (\tr\bC)^{2} - \tr(\bC^2)  \big).
    \label{SecondInvariant}
\end{align}
\end{subequations}
The first Piola--Kirchhoff stress tensor for a compressible Mooney--Rivlin solid is
\begin{equation}
    \Pel =J^{-2/3}(\mu_{1} 
    +\mu_{2}\bar{I}_{1}) \bF 
    - J^{-4/3} \mu_{2}\bF \bC +(k \log J 
    - \frac{1}{3}\mu_{1}\bar{I}_{1} 
    - \tfrac{2}{3}\mu_{2}\bar{I}_{2} )\FInvT
\label{MRConstitutiveStress}
\end{equation}

The fluid phase is assumed to obeys the standard Darcy's law in the spatial configuration
\begin{equation}
    \bq= - \bk(x,t) \, \grad p,
    \label{DarcyLawSpatial}
\end{equation}
where $\bk(x,t)$  is the permeability tensor and $p$ the fluid pressure. The tensor $\bk(x,t)$ is usually assumed to be symmetric and positive-definite. For isotropic porous media, a simple constitutive characterization is 
\begin{equation}
    \bk(x,t) = k(x,t) \boldsymbol{I} \, ,
    \label{PermeabilityTensor1}
\end{equation}
where $k(x,t)$ is a scalar function. This permeability tensor, spherical and with principal axes not affected by the deformation, is said to be unconditionally isotropic. 
Exploiting the properties of isotropy, a general result for the spatial representation of $\bk$ was presented in \cite{ateshian:2010}
\begin{equation}
    \bk_{\rm{AW}}(x,t) = k_{0}(x,t)\boldsymbol{I} + k_{1}(x,t)\boldsymbol{b} + k_{2}(x,t)\boldsymbol{b}^{2} \, ,
    \label{PermeabilityTensorAW1}
\end{equation}
where $\boldsymbol{b} = \bF \cdot \FT$ is the left Cauchy--Green deformation tensor. In the reference configuration, it reads
\begin{equation}
    \bK_{\rm{AW}} = k_{0}\CInv + k_{1} \bI + k_{2}\bC \, .
    \label{PermeabilityTensorAW2}
\end{equation}
The referential permeability tensor is obtained as the pull-back of $\bk(x,t)$
\begin{equation}
    \bK(\bX,t) = J \FInv \bk(x,t) \FInvT \, .
    \label{PermeabilityTensor2}
\end{equation}
The referential form of the Darcy's law is
\begin{equation}
    \bQ = J \FInv \bq= - J \FInv
    \bk(x,t) \, \grad p = - 
    \bK(\bX,t) \, \Grad p \, .
    \label{DarcyLawReferential}
\end{equation}

\section{Weak form of balance equations and Finite Element discretization}
\label{sec:FiniteElements}

To derive the finite element form of the balance equations, we begin by inserting the Darcy law into the continuity equation \eqref{EquationContinuity}, obtaining
\begin{subequations}
\begin{align}
    & \Div \left (\bK \Grad p \right) - \dot{J} = 0 && \rm{in} \, \Bref,
    \label{EquationContinuity2} \\
    & \Div(\Phis \Pel - J p \FInvT) = \bO && \rm{in} \, \Bref \, .
    \label{EquationEquilibrium2} 
\end{align}
\end{subequations}
Next, we multiply the strong form of the continuity equation by a scalar field $\eta$ (test function) and integrate over the domain
\begin{equation}
    \int_\Bref { \Div (\bK \Grad p ) \, \eta} \, dV - \int_\Bref {\frac{\partial J}{\partial t} \, \eta} \, dV = 0 \, .
    \label{continuityWeak1}
\end{equation}
Integration by part and an application the divergence theorem,  Eq.~\eqref{eq:Appendix3}, lead to
\begin{equation}
    \int_\Bref { \bK \Grad p \cdot \Grad \eta } \, dV  -   
    \int_\Bref { J \Grad \Vs : \FInv  \, \eta} \, dV
    =
    \int_{\partial_{\rm f} \Bref^{
    \rm q}} { \bar \bQ \, \eta } \, dA \, ,
    \label{continuityWeak2}
\end{equation}
where the boundary integral is non zero only on the Neumann boundary, where it has to satisfy the condition \eqref{BoundaryNeumannFluid}. 
We repeat the same procedure for the linear momentum balance, using a vector test field $\bxi$, and obtain
\begin{equation}
    \int_\Bref {\Div \left( \Phis \Pel - J p \FInvT \right) \cdot \bxi} \, dV
    =
    0 \,  .
    \label{linearMomentumWeak1}
\end{equation}
Integration by part and an application the divergence theorem lead to
\begin{equation}
    - \int_\Bref { J p \FInvT : \Grad \bxi } \, dV 
    + 
    \int_\Bref {\Phis \Pel : \Grad \bxi } \, dV  = \int_{\partial_{\rm s} \Bref^{\rm t}} { \bar \bT \cdot \bxi } \, dA  \, .
    \label{linearMomentumWeak2}
\end{equation}
Inserting a finite element discretization in the weak balance equations, as detailed in Appendix B, we obtain two coupled systems of equations
\begin{align}
    &  \bM \, \bP(t)  +  
    \bH \, \dot{\bU} (t)
    =
    \bB(t) \label{finiteElementP} \\
    & \bH^{\rm T} \bP(t)
    + 
    \bR^{\rm int}(\bU, t) 
    = \bR^{\rm ext} (t) \, .
    \label{finiteElementD}
\end{align}
Upon time discretization on the assigned linear loading history into time steps $\Delta t_n = t_{n+1}-t_n$, the solution of the nonlinear system is achieved using a standard monolithic Newton-Raphson approach\footnote{The numerical algorithm is implemented in the MOOSE software ecosystem \cite{harbour2025moose}}, approximating the rate of the displacement in Eq.~\eqref{finiteElementP} as 
\begin{equation}
    \dot{\bU}(t_{n+1}) \approx \frac{\bU_{n+1} - \bU_n}{t_{n+1}-t_n} \, .
    \label{rateApproximation}
\end{equation}
where $\bU_n$ is the displacement solution obtained at time $t_n$. 

\section{Numerical model of the cornea}
\label{sec:numericalmodelling}
The cornea is anchored to the surrounding eye tissues at the limbus, a stiffer region that connects the transparent cornea to the white sclera and to the colored iris. Together, the iris and cornea define the boundaries of the anterior chamber of the eye, which is filled with aqueous humor maintained at a physiological pressure of about 15–18~mmHg. This fluid, rich in ions, interacts with the stroma through a pump-leak mechanism regulated by the endothelium. In healthy conditions, the pump-leak mechanism provides a zero net-flux of fluids.

\subsection{Geometry}
\label{ssec:geometry}

\begin{figure}[H]
    \centering
    \includegraphics[width=\linewidth]{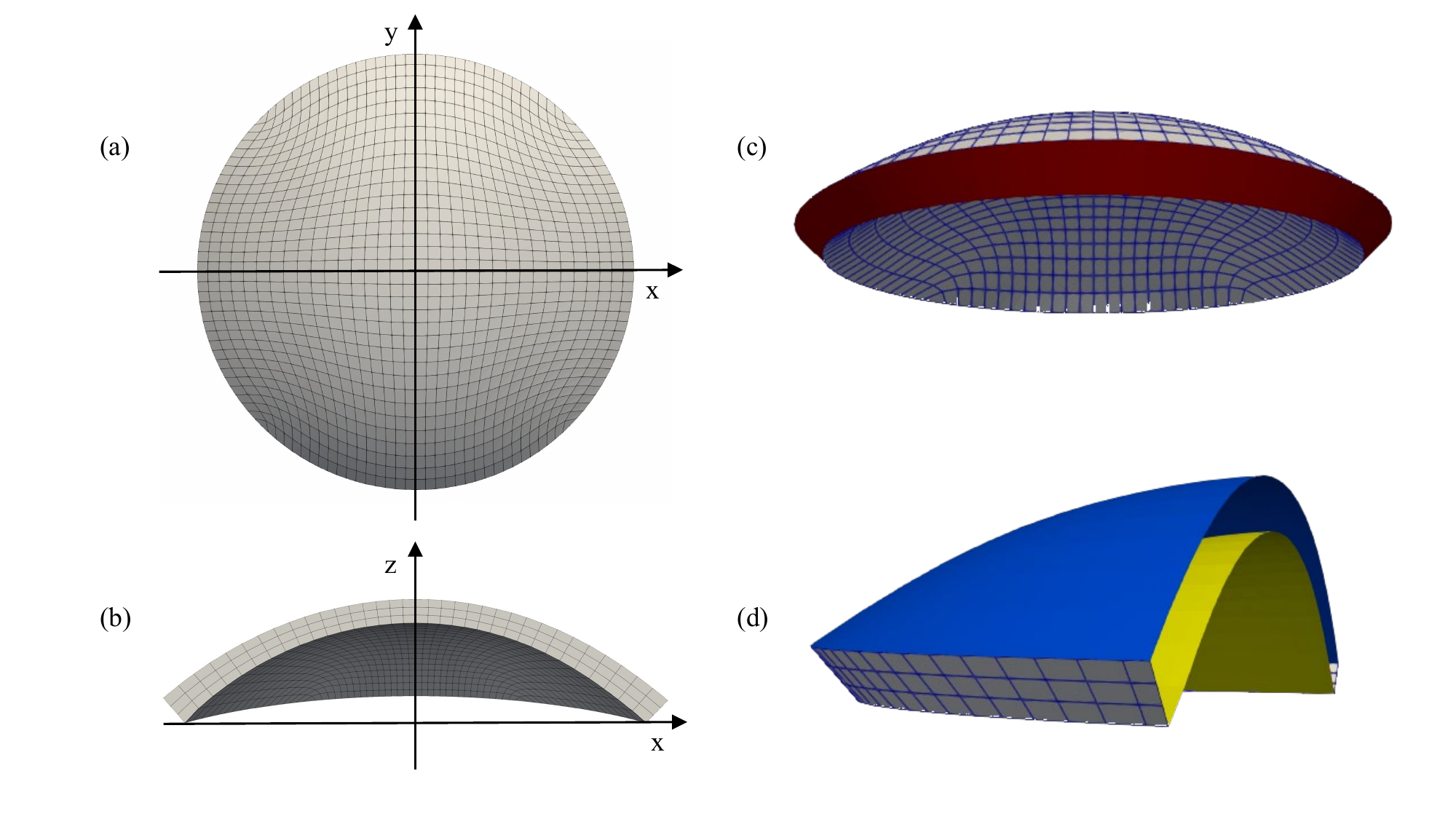}
    \caption{\footnotesize Finite element model of the patient specific cornea used in this study. (a) View of the anterior surface with the surface discretization, comprising 2500 nodes and 1728 hexahedral solid elements. (b) Nasal-Temporal meridional section, with the thickness discretization. (c) In red, limbial surface where Dirichlet type boundary conditions are applied: fixed displacements on $\partial_{\rm s} \Bref^{\rm u}$, and zero fluid pressure on $\partial_{\rm f} \Bref^{\rm p}$. (d) In yellow, posterior surface where Neumann type boundary conditions are applied: variable IOP on $\partial_{\rm s} \Bref^{\rm t}$, and variable fluid flux on $\partial_{\rm f} \Bref^{\rm q}$. In blue, anterior surface where Neumann type boundary conditions are applied: no stress conditions on $\partial_{\rm s} \Bref^{\rm t}$, and no flux conditions on $\partial_{\rm f} \Bref^{\rm q}$. }
    \label{fig:fig2}
\end{figure}

A patient specific geometry of the cornea and the corresponding finite element discretization into 8-noded brick elements are generated using a solid modeling tool specifically developed for human corneas. Details on the solid modelleer and mesh generator are available elsewhere \cite{simonini:2015,montanino:2023}. The dome-shape geometry of the cornea is acquired from digitalized diagnostic images of the physiological configuration of a human cornea. The discretization into finite elements has been chosen to follow in the central part of the cornea an orthogonal grid smoothly rotating to reach a radial-circumferential orientation of the elements at the limbus. The central grid is aligned with the two main meridians of the cornea, e.~g., the nasal-temporal (NT) an the superior-inferior (SI). The optic axis, orthogonal to NT and SI, crosses the cornea in correspondence of the center, and, for healthy corneas, defines the apex location, see Fig.~\ref{fig:fig2}(a,b).

Note that the physiological configuration is a spatial configuration characterized by a displacement field. Thus, as explained in \cite{pandolfi:2006}, a correct stress analysis requires the identification of the unstressed geometry (that can be considered as reference configuration), which depends on the chosen material model and on the value of the IOP. Nevertheless, in this study we omit the identification of the unstressed configuration, since we are focused on other aspects of the modeling and we are not conducting  simulations with predictive finality. 

\subsection{Boundary conditions}
\label{ssed:boundaryConditions}

In the present study, we do not model explicitly the endothelium at the posterior surface of the cornea, but a we apply boundary conditions corresponding to the action of an endothelial pump-leak mechanism in place to preserve the fluid content of the stroma. The overall boundary conditions for the biphasic cornea model are therefore chosen as follows.

The limbus is considered as a Dirichlet-type boundary for both static and hydraulic problems. Displacement components and fluid pressure of all the boundary nodes are set to zero, Fig.~\ref{fig:fig2}(c). The anterior and posterior surfaces of the cornea are set as Neumann boundaries. On the anterior surface, surface tractions and fluid fluxes are set to zero, Fig.~\ref{fig:fig2}(d). On the posterior surface, tractions define a normal pressure, with intensity corresponding to the intraocular pressure growing from zero to $p_{\rm IOP} = 18$~mmHg, and a net fluid flux $\zeta$ reproduces the pump-leak mechanism \cite{vanone:2025}, as
\begin{equation}
    (\Ps - J p \FInvT) \bN = - p_{\rm IOP} \bN
    \quad {\rm on} \; \partial_{\rm s} \Bref^{\rm t}
    \,, \quad
    \bQ \cdot \bN = \zeta 
    \quad {\rm on} \; \partial_{\rm f}\Bref^{\rm q}
    \label{boundaryConditions}
\end{equation}
In physiological state, we hypothesize that the net fluid flux is null, i.e. $\zeta = 0$, since the intake of fluid due to the swelling pressure of the cornea is balanced by the pump-leak mechanism of the endothelium that involves the transport of ions \cite{vanone:2025}. In pathologic conditions, the flux is set to $10^{-5}$ mm/s.

\subsection{Modeling damage}
\label{ssec:damage}

The physiological state of the human cornea is characterized by a balanced transport of fluid and ions that preserves the homeostasis of the tissue. With a suitable choice of boundary conditions, corresponding to a steady state filtration of the fluids, the model is able to describe the physiological strain and stress distributions, with negligible differences with respect to a purely mechanical model. Nevertheless, the primary motivation of the introduction of a solid-fluid mixture for the stroma is to capture degenerative states of the anterior eye that lead to the loss of refractive functions, in particular ectasia and corneal swelling. Traditional monophase continuum models comply with the incompressibility condition, and fail to capture the main features of the pathology, particularly in terms of the extent of corneal thinning and the regional variability in thickness, typically observed in clinical observations.

\begin{figure}[H]
    \centering
    \includegraphics[width=0.95\linewidth]{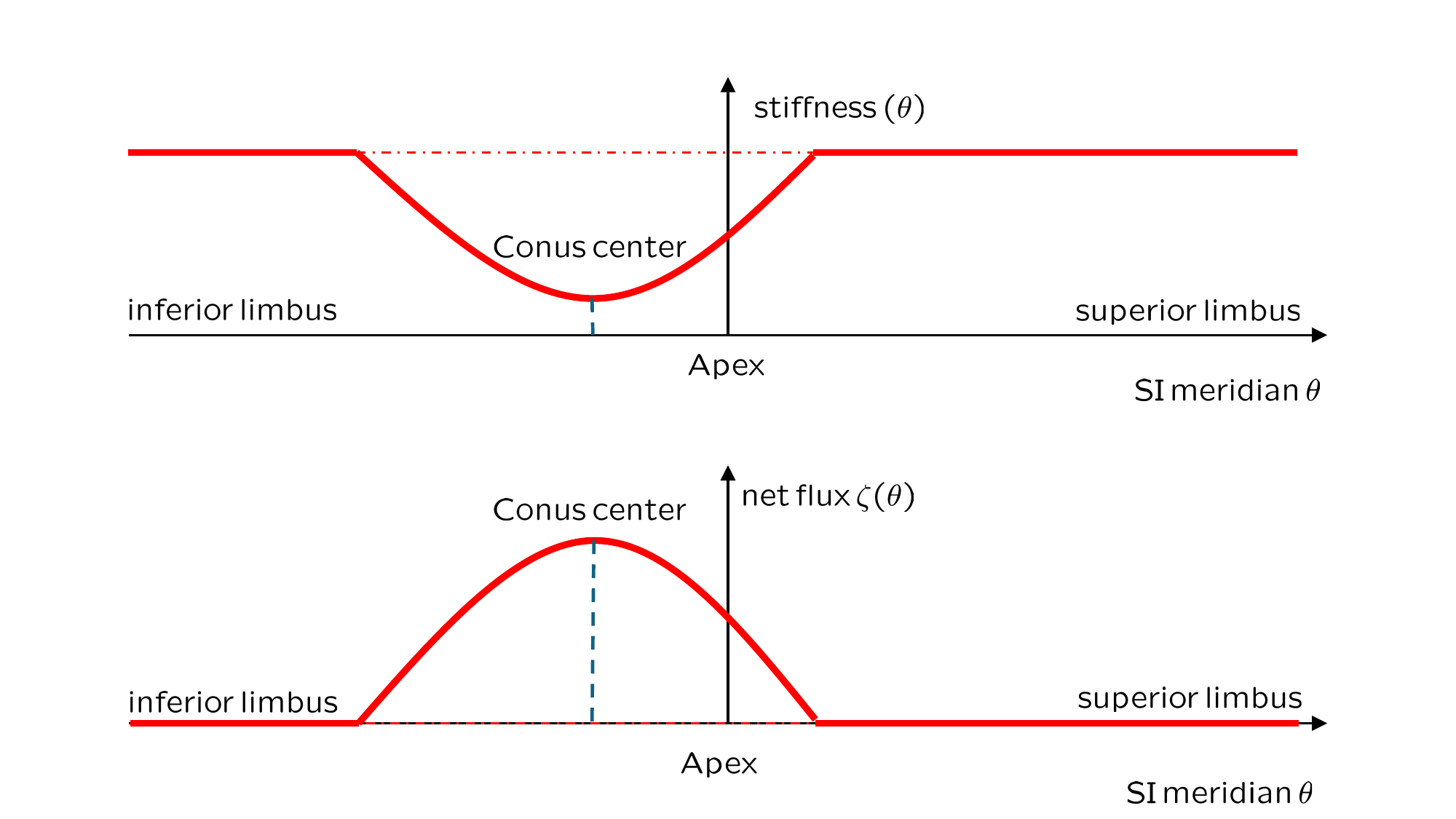}
    \caption{\footnotesize Assumed quadratic profiles of mechanical and hydraulic degenerations due to keratoconus. Plots show the distribution of the material stiffness parameters and of the net outflux within a circular area around the conus center, located at an offset form the corneal apex along the SI meridian.}
    \label{fig:fig3}
\end{figure}

The numerical modelling of corneal pathologies leading to corneal swelling is a complex process that requires the introduction of an equivalent endothelium layer, see, e.g., the approach described in \cite{vanone:2025}, which will be addressed in an incoming work. In the present study we explore the possibility to capture the modification of curvature and thinning observed in ectasia and keratoconus. Taking a purely phenomenological approach, we follow here the idea presented in \cite{pandolfi:2019} and simply describe the effects of keratoconus by reducing selectively the mechanical properties (i.e., any stiffness parameter of the model) over a limited area of the cornea. We assume that the deterioration is described by a damage mapping with circular shape and quadratic distribution, null at the boundary of the circular area and maximum at the center of the conus. In addition, we assume that the pathology alters the delicate pump-leak equilibrium in the same circular area, causing a non-zero outflux $\zeta$ with quadratic distribution. Fig.~\ref{fig:fig3} shows the profile of the stiffness and of the outflux along the SI meridian. 

\section{Results of the numerical simulations}
\label{sec:Results}

The biphasic model for the biological material has been employed to investigate the possibility to capture unexplored aspects of physiological and pathological behaviors of the human cornea. In this regard, we want to verify if the inclusion of non-zero flux boundary conditions on the posterior surface of the cornea is able to capture the change of the geometry observed during the evolution of the keratoconus. 

In consideration of the simplicity of the solid model chosen for this study, where the anisotropy due to the presence of collagen fibers has been disregarded, we confine our investigation to the assessment of the influence of only two significant parameters of the model. Thus, we explore the relevance of the equivalent bulk modulus $k$ that assesses the incompressibility of the solid phase, and of the maximum value of the net flux $\zeta$ applied to the posterior surface of the cornea. For each parameter we select two limiting values that define a meaningful range of applicability of the model. The values of the {\it fixed} mechanical parameters of the solid are derived from previous studies, as representative of healthy behavior of the tissue \cite{miller:2025}.

We refer to two healthy cases, Case 1 and Case 2, which are somehow representative of our previous purely mechanical models of cornea in healthy conditions. Case 2 differs from Case 1 for a reduced value of the bulk modulus from $k=5$ to $k=2.5$ MPa. The third and the fourth models (Cases 3 and 4) implement the reduction of the mechanical stiffness and the inclusion of a non-zero net flux on the posterior surface of the cornea, according to the profiles shown in Fig.~\ref{fig:fig3}. 
The mechanical and hydraulic parameters adopted in this study are collected in Table~\ref{tab:parameters}. We remark once again that the model is not proper for a validation against clinical data on keratoconus, given the lack of the collagen fiber reinforcement, but it is sufficient to capture the thinning and flattening of the corneal shell.

The numerical simulations reproduce quasi-static inflation tests (i.e., ex-vivo posterior surface pressurization tests with fixed limbus), where the pressure is progressively increased from zero to 18~mmHg, representative of the average physiological IOP. When the non-zero flux is active, it is also progressively increased from zero to its final value.

\begin{table}[H]
    \centering
    \begin{tabular}{cccccc}
         \hline
         Parameter & Case 1 & Case 2 & Case 3 & Case 4 & Units\\
         \hline
         $\Phi_{\rm{s}}$ & 0.2 & 0.2 & 0.2 & 0.2 &  - \\
         $k_{0}$  &  1.5 & 1.5 & 1.5 & 1.5 & $10^{-3} \, \rm{mm}^{4}/\rm{Ns}$ \\
         $\mu_{1}$ & 1 & 1 & 1 & 1 & $\rm{MPa}$ \\
         $\mu_{2}$ & -0.35 & -0.35 & -0.35 & -0.35 & $\rm{MPa}$\\
         $k$ & 5 &  2.5 &  5 &  2.5 & $\rm{MPa}$\\
         $\zeta$ & 0 & 0 &  $10^{-5}$ & $10^{-5}$ & mm/s \\
         \hline
    \end{tabular}
    \caption{\footnotesize Values of the material parameters used in the numerical simulations. Case 1: Healthy cornea, with reference material values and $k=5$ MPa. Case 2: Healthy cornea, with reduced bulk modulus $k=2.5$ MPa. Case 3: Diseased cornea, with $k=5$ MPa. Case 4: Diseased cornea, with reduced bulk modulus $k=2.5$ MPa. Cases 3 and 4 include the variability of stiffness and posterior surface outflux as shown in Fig.~\ref{fig:fig3}. }
    \label{tab:parameters}
\end{table}

We begin with showcasing the ability of the biphasic model to capture a marked thinning effect similar to the one observed in the keratoconus. 

\begin{figure}[H]
    \includegraphics[trim=50 0 260 0,clip,width = \textwidth]{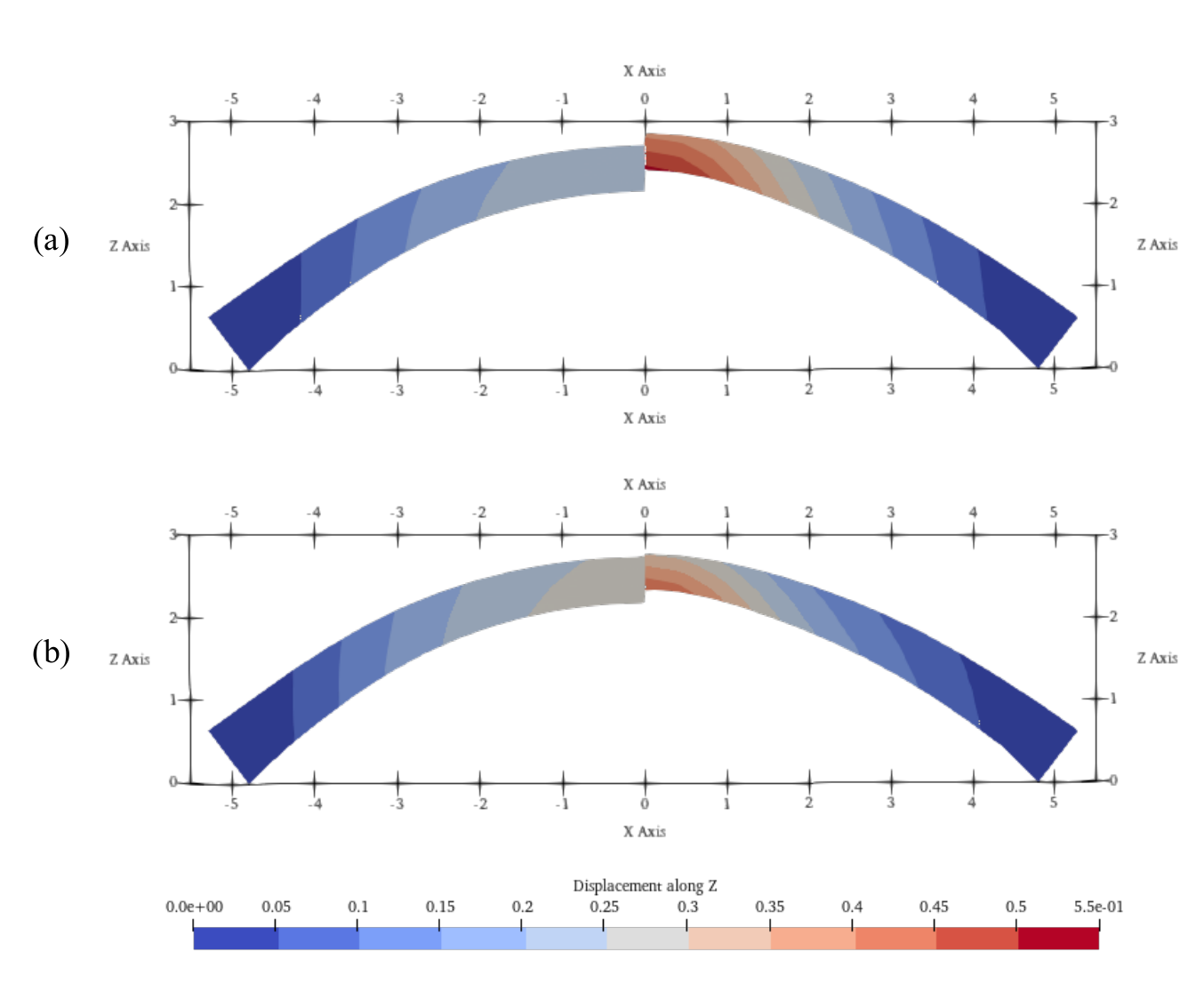}
    \caption{Elevation maps (component of the displacements in the direction of the optic axis) computed with the models of the human cornea, under the action of the physiological IOP, 18 mmHg. Plots refer to the NT meridional section. The figure compares the results for healthy and keratoconus corneas. The diseased corneas are characterized by a reduction of the mechanical stiffness of the tissues and by a non-zero flux from the endothelial side. The two images refer to a different choice of the bulk modulus: (a) $k=5$ MPa, (b) $k=2.5$ MPa.} 
    \label{fig:fig4}
\end{figure}
\begin{figure}[H]
    \includegraphics[trim=0 0 0 0,clip,width = \textwidth]{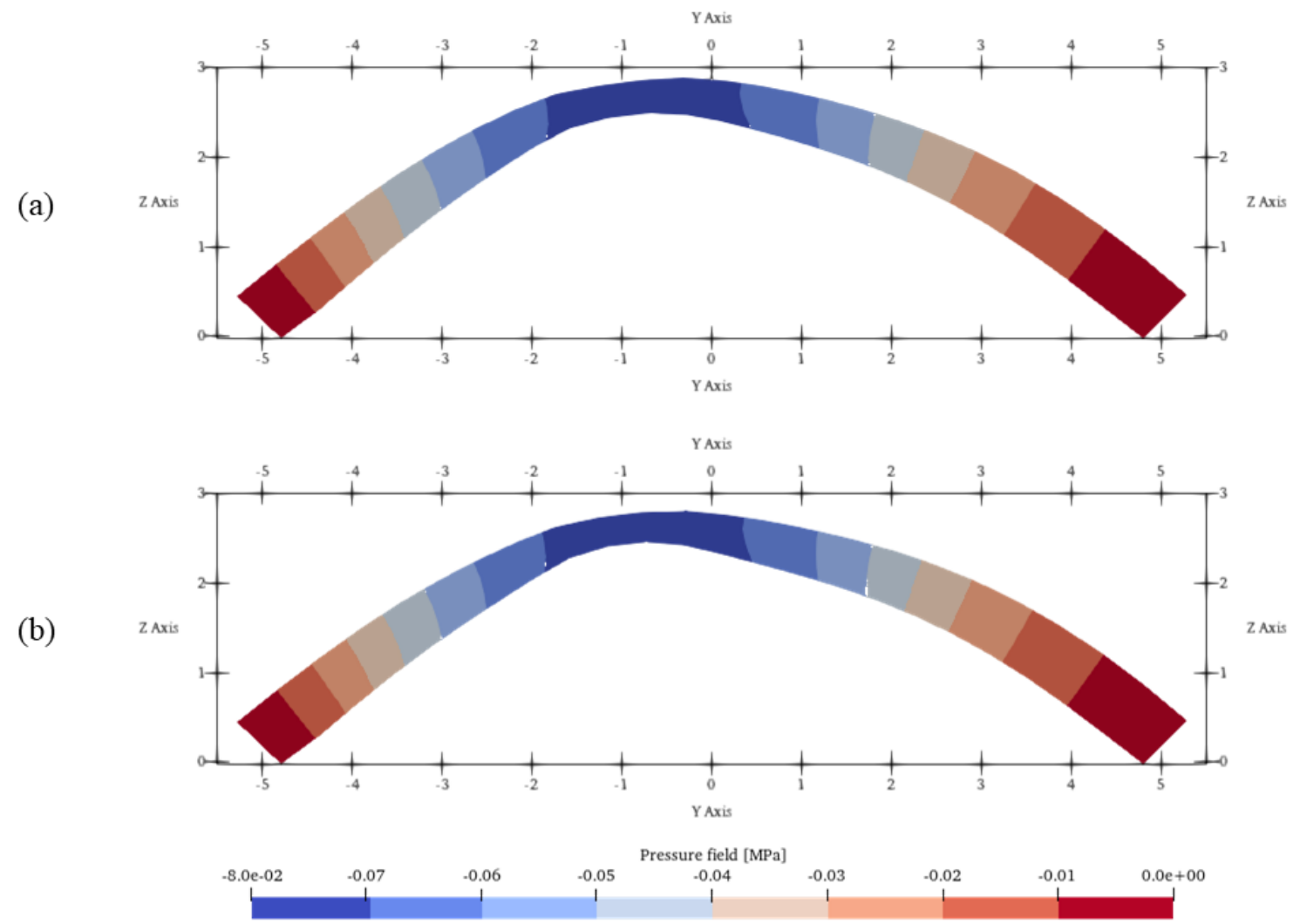}
    \caption{Fluid pressure maps $p$, computed with the biphasic models, of the human cornea under the action of the physiological IOP, 18 mmHg. Plot refers to the SI meridional section. The figure compares the results for Case 3 and Case 4 in Table \ref{tab:parameters}. The diseased cornea are characterized by a reduction of the mechanical stiffness of the tissues and by a non-zero flux from the endothelial side. The two images refer to a different choice of the bulk modulus:  (a) $k=5$ MPa, (b) $k=2.5$ MPa.} 
    \label{fig:fig5}
\end{figure}

Fig.~\ref{fig:fig4} compares NT meridional half-sections (in the direction of the $x$ axis) of the cornea stressed by the physiological IOP, for healthy and keratoconus cases. Color maps refer to the displacement component in the direction of the optic axis ($z$ axis).
Note that the most damaged area of the keratoconus is not visualized becasue it is located below the NT axis. More in detail,
Fig.~\ref{fig:fig4}(a) compares the healthy Case 1 (left side) with the keratoconus Case 3 (right side), situations characterized by a high value of the bulk modulus, $k=5$~MPa. Similarly, Fig.~\ref{fig:fig4}(b) compares the healthy Case 2 (left side) with the keratoconus Case 4 (right side), characterized by a low value of the bulk modulus, $k=2.5$~MPa.  
Evidently, the biphasic models predict a marked thinning of the tissue in the damaged region. The displacements for the diseased case in Fig.~\ref{fig:fig4}(a) reaches higher values than the ones in Fig.~\ref{fig:fig4}(b) due to the interplay between a reduced bulk modulus and the non-zero flux $\zeta$. 
Note that the situation shown in Fig.~\ref{fig:fig4}(b), where the anterior displacement at the apex in the healthy and diseased cases are similar, is a fortuitous  consequence of the parameter choice and not a generally expected result. Furthermore, the images do not show the maximum thinning, which is reached at the center of the cone and can be appreciated in Fig.~\ref{fig:fig5}.

Fig.~\ref{fig:fig5} compares the pressure fields along the SI meridian, at the physiological IOP, of the deseased models (Case 3 and Case 4 in Table~\ref{tab:parameters}). The enforcement of a positive outflux $\zeta$ combined with a zero pressure at the limbus produces significant pressure gradients, with compressive pressure peaks of $-0.08$~MPa. The pressure field gradually increases from the center of the conus to the limbus. The images clearly show the thickness reduction at the center of the conus, which reaches about 42\% of the original thickness.

\begin{figure}[H]
\vspace{-0.3cm}
    \centering
    \includegraphics[width=0.85\linewidth]{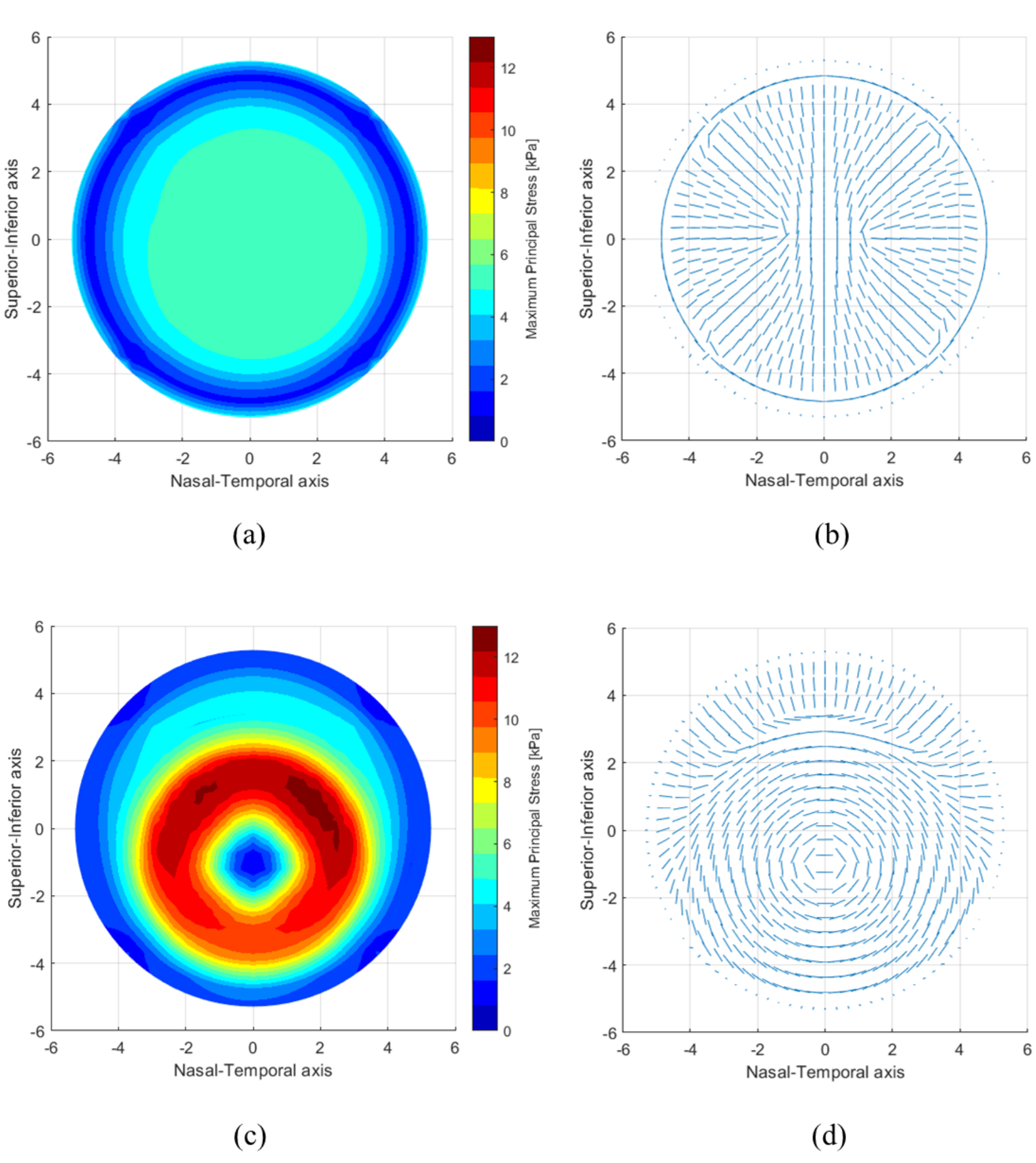}
    \caption{\footnotesize Distribution of the maximum principal deviatoric stress for the healthy cornea (up, Case 1) and diseased cornea (down, Case 3), for a high value of the bulk modulus, $k=5$ MPa. Left: maps of the maximum value. Right: maps of the principal directions.}
    \label{fig:fig6}
\end{figure}
\begin{figure}[H]
\vspace{-0.3cm}
    \centering
    \includegraphics[width=0.85\linewidth]{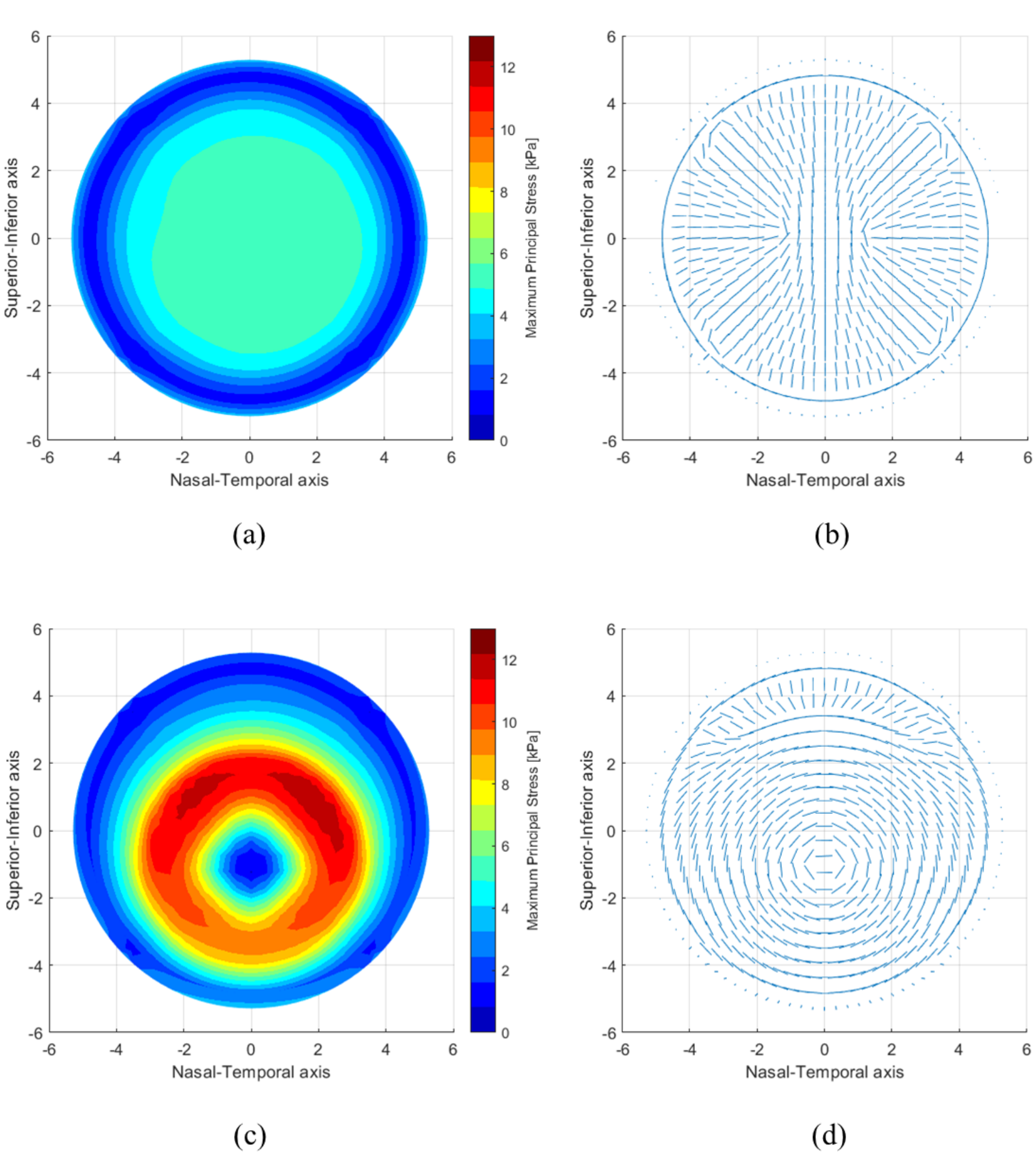}
    \caption{\footnotesize Distribution of the maximum principal deviatoric stress for the healthy cornea (up, Case 2) and diseased cornea (down, Case 4), for a reduced value of the bulk modulus, $k=2.5$ MPa. Left: maps of the maximum value. Right: maps of the principal directions.}
    \label{fig:fig7}
\end{figure}
Figs.~\ref{fig:fig6}-\ref{fig:fig7} compare the maps of stresses on the anterior surface of the cornea at the physiological IOP, revealing important differences between healthy and diseased tissues. The images show the maps of the maximum principal value of the deviatoric Cauchy stress of the solid $\bs_{\rm s}$, defined as
$$\bs_{\rm s} = \bsigmas - \frac{1}{3}~\rm{tr}~\bsigmas \,,$$
and the maps of the corresponding principal directions (tensile isostatics), for the two chosen values of the bulk modulus, $k=5$ MPa, Fig.~\ref{fig:fig6}, and $k=2.5$ MPa, Fig.~\ref{fig:fig7}. Healthy and diseased tissues are considered. Remarkably, in the healthy cases, shown in Figs.~\ref{fig:fig6}-\ref{fig:fig7}(a), the maximum deviatoric Cauchy stress attains a uniform distribution, demonstrating the optimal engagement of the tissue typical of homeostatic conditions. By the way of contrast, in the diseased cases, shown in Figs.~\ref{fig:fig6}-\ref{fig:fig7}(c), the maximum stress reaches very high values within the circular annulus surrounding the conus center.

\begin{figure}[H]
    \centering
    \includegraphics[width = 0.8\textwidth]{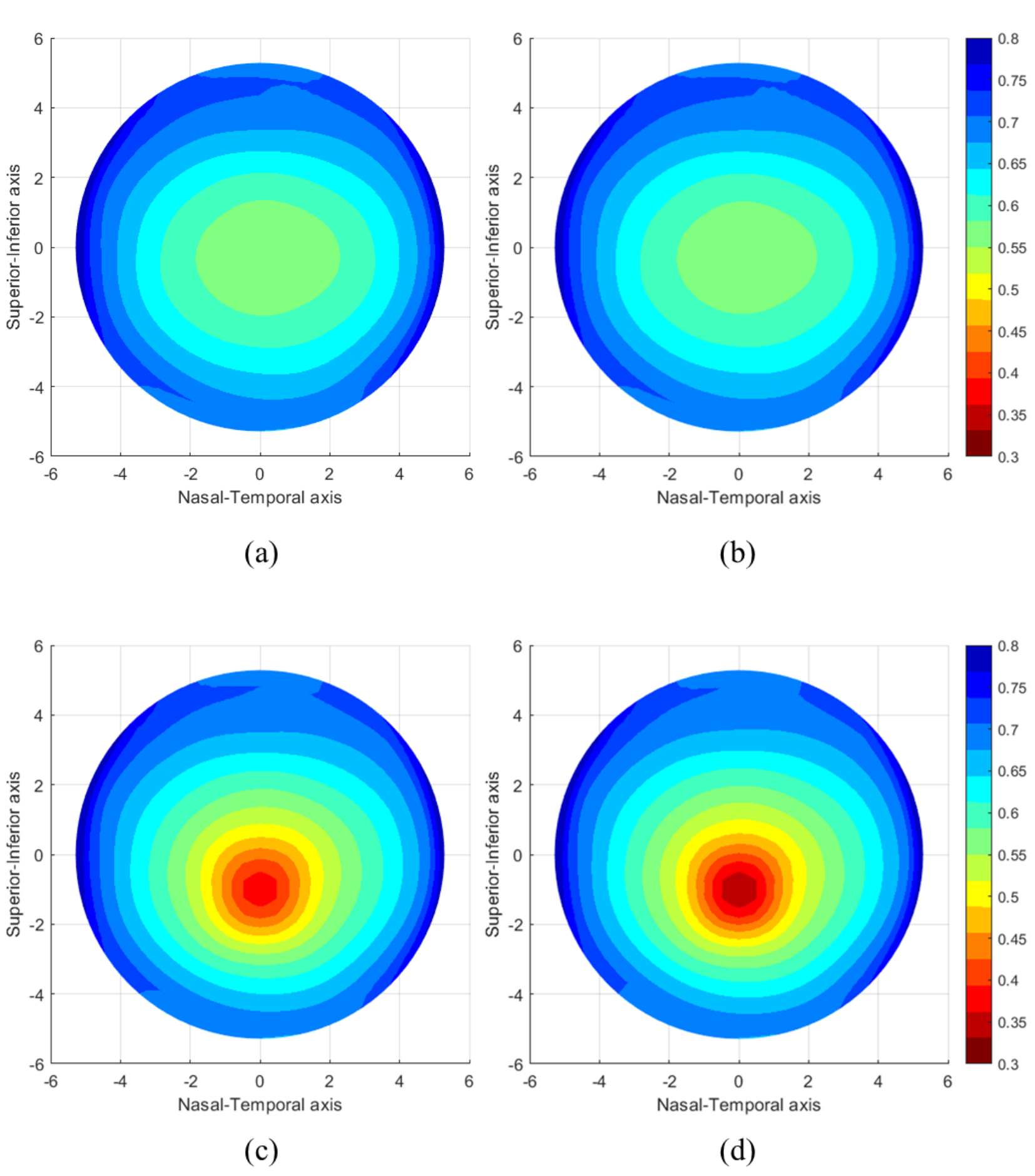}
    \caption{Pachymetry maps computed with the models of the human cornea, under the action of the physiological IOP, 18 mmHg. Results are shown for healthy corneas: (a) Case 1, (b) Case 2, and diseased corneas: (c) Case 3, and (d) Case 4.}
\label{fig:fig8}
\end{figure}
The redistribution of the stresses piling up in less damaged areas is a typical behavior of tissues to compensate a localized strong reduction of the stiffness.
Within the limits of unavoidable artifacts induced by the adopted discretization, in the healthy cases the tensile isostatics follow the meridians of the shell, see Figs.~\ref{fig:fig6}-\ref{fig:fig7}(b), revealing the behavior expected in a pressurized spherical dome. 
In the keratoconus cases, the tensile isostatics are parallel circumferences centered at the cone, see Figs.~\ref{fig:fig6}-\ref{fig:fig7}(d), revealing the behavior expected in a pressurized cone. These results are in agreement with the analysis of stresses in healthy and ectatic eyes documented in \cite{angelillo:2020} and with classical results in linearized shell theory.

Fig.~\ref{fig:fig8} shows the computed pachymetry (thickness) maps, projected onto the anterior surface of the cornea. In the clinical practice, pachymetry maps are particularly useful to follow the  evolution of keratoconus and to localize the most damaged areas. The images show the maps at the physiological IOP, for Case 1 and Case 2, Figs.~\ref{fig:fig8}(a) and (b), and the diseased cases, where the mechanical damage and the posterior outfow are considered, Figs.~\ref{fig:fig8}(c)-(d). Keratoconic maps clearly reveal tissue thinning at the conus, clearly visible in section in Fig.~\ref{fig:fig4}. The maximum value of thickness reduction reaches 42\% of the original thickness, which is a result hard to achieve with purely mechanical models.

\begin{figure}[H]
    \centering
    \includegraphics[trim=0 0 0 0,clip,width = 0.8\textwidth]{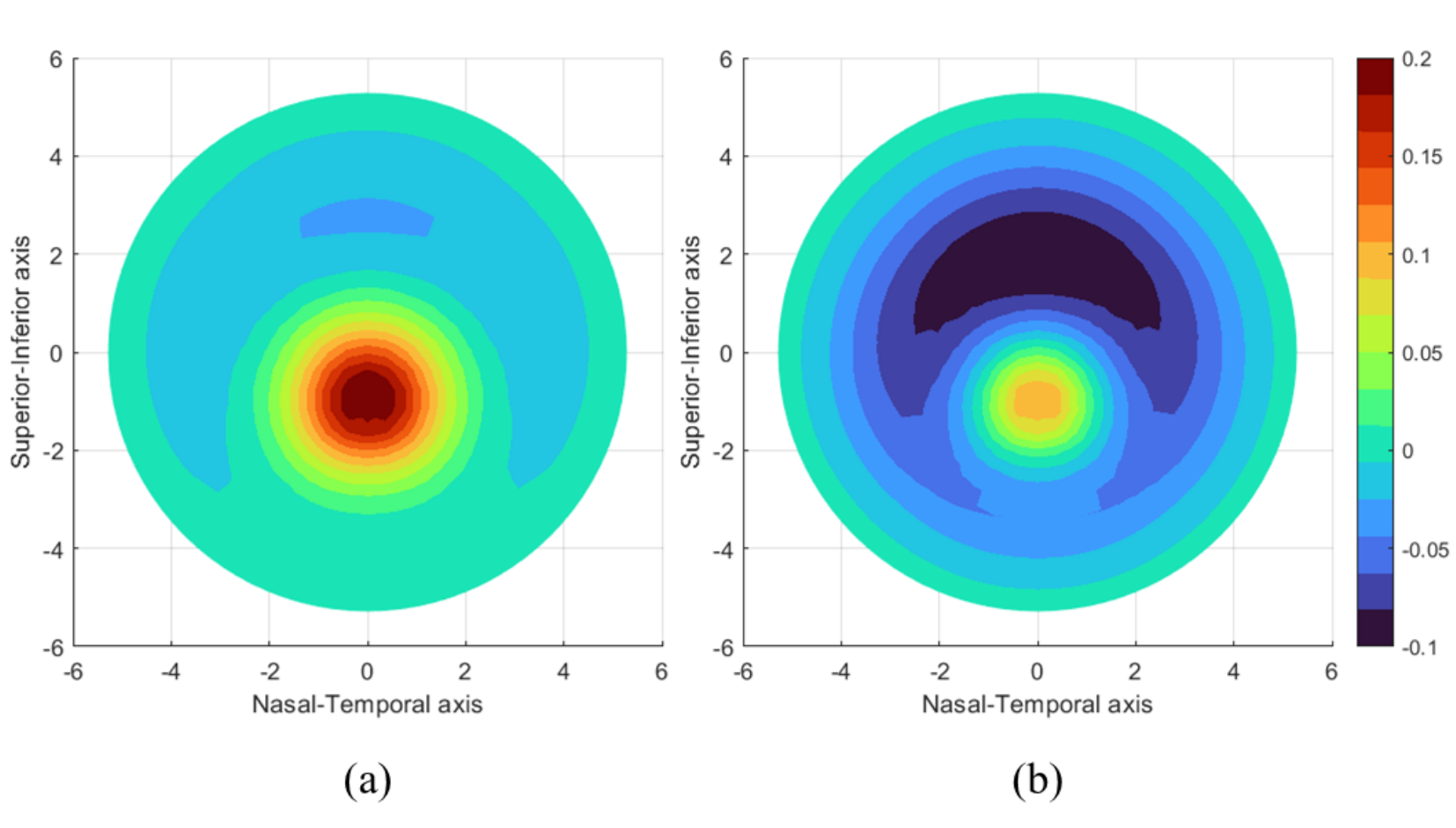}
    \caption{Relative altimetry maps computed with the models of the human cornea, under the action of the physiological IOP, 18 mmHg. Maps visualize the difference in the relative displacement of the anterior surface of the cornea between the diseased corneas (Case 3 and Case 4) and the corresponding healthy ones (Case 3 and Case 4). The cornea with a high value of $k$ (a) is less affected by the loss of fluid through the endothelium than the cornea with a lower value (b).}
\label{fig:fig9}
\end{figure}
To better understand the clinical relevance of adopting a biphasic model to describe the damage induced by keratoconus, we make recourse to relative altimetry maps at physiological IOP for the anterior surfaces, see Fig.~\ref{fig:fig9}. These maps are constructed by subtracting the elevation maps of healthy corneas from the ones of damaged corneas. The images, in order, show the relative elevation maps for the cases $k=5$ MPa (a) and $k=2.5$ MPa (b). The images show an important geometrical aspect, that is often significant to the diagnosis of keratoconus. In both cases, the superior half cornea is characterized by a retraction (backward motion) with respect to the healthy case, indicating that the curvature of that zone is reducing, and the shell switches from a spherical shape to a more conical shape. Again, these results are in agreement with the analysis of curvature in healthy and ectatic eyes documented in \cite{angelillo:2020}.

\section{Conclusions}
\label{sec:Conclusions}

We presented a new approach to the biomechanics of the human cornea that employes a biphasic model to describe physiological and pathological states of the tissue.In this first study, we made a few drastic assumptions.

First, we considered for the solid part an isotropic behavior which clearly contradicts the well-known anisotropic  micro-architecture of the stroma. The reason of this choice is the need to simplify the implementation of the coupled model into a finite element code, to assess the relevance of a biphasic model in describing pathological behaviors. 
Second, although the study of the configurational progression of the keratoconus is one of out future objectives, we did not make any assumptions of growth or remodeling for the solid phase, since the effective reduction of mechanical properties (i.e., stiffness) of the tissue can be sufficiently well explained in the long term by a phenomenological damage variable distribution, as already proved in recent studies \cite{miller:2025}.

Since the focus of the study is a preliminary exploration of the effects of hydraulic-mechanical coupling in a model of the human cornea, we are not showing here the results of  parametric studies conducted on the fixed parameters of the model, which for the simplified model resulted to be marginal and lacking of physical significance.
We also omit the results of convergence analysis on the mesh size, because we are not pursuing realistic prediction of the evolution of a disease. 

A comment is worth on the choice of fixed boundary conditions for the solid phase. As it was done in our previous studies, the constraints at limbus should  mimic the presence of compliant surrounding tissues, that are not explicitly modeled. Here we applied fixed displacement boundary conditions, while more compliant boundaries, that avoid the insurgence of bending stresses at the limbus, shall instead be used for a more realistic description of the cornea in-vivo. The so-called rotating boundaries, described in \cite{Pandolfi2008}, can be easily introduced for analyzing patient specific cases.

Finally, the finite kinematics adopted in this study calls for the identification of the ideal unstressed configuration, from where the numerical analysis should be conducted. This feature will be considered for more realistic patient specific simulations \cite{montanino:2018}.

We are aware that simplifications used here hamper the predictive potential of the model. Nevertheless, from this study we have learned that, in pathological conditions, the biphasic nature of the stroma should not be neglected, since it represents an important property of living tissues, where growth or remodeling are phenomena that do not appear of major relevance.

The next steps toward the construction of a predictive biphasic model of the stroma include incorporating collagen fibers, responsible of anisotropy, and including a posterior layer of tissue (or an equivalent membrane) representative of the processes taking place in the endothelium, such as the pump-leak function and the swelling control.

\section*{Acknowledgments}
This work was supported by the Italian Ministry of University and Research (MUR) under the PRIN 2022 project CORTIS (Grant No. 2022TWKA72). We acknowledge the support of the Italian National Group of Physics-Mathematics (GNFM) and the Italian National Institution of High Mathematics ``Francesco Severi'' (INDAM).

\section*{Appendix A. Mathematical relationships}
\label{sec:appendixA}

We recall here the mathematical relationships necessary to obtain governing equations.
\\
The material derivative with respect to the motion of the phase $\alpha$ is 
\begin{equation}
    \rm{D}_{\alpha} (\bullet) = \partial_{t} (\bullet) + \grad (\bullet) \cdot \va \, .
    \label{eq:Appendix1}
\end{equation}
The rate of the Jacobian $J$ in spatial and referential forms is given by
\begin{equation} 
    \dot J = J \ddiv \vs = J \,\tr ( \Grad \Vs \, \FInv ) = J \, \Grad \Vs : \FInvT \,.
    \label{eq:Appendix3}
\end{equation}
The referential percentage volume of the solid phase is 
$$\Phi_{\rm{s}} = J [\phi_{\rm{s}} \circ (\bchi(X,t),t)] \, ,$$ 
and of the fluid phase is
$$\Phi_{\rm{f}} = J [\phi_{\rm{f}} \circ (\bchi(X,t),t)] \, .$$ 
The gradient and the divergence operators in the reference configuration are linked to the spatial counterpart as
$$ \Grad ( (\bullet) \circ \bchi(X,t) ) = \big(\grad (\bullet) \, \bF \big) \circ \bchi(X,t) \, , $$
$$\Div(J\FInv (\bullet) ) = J \, \ddiv (\bullet) \, ,$$
for any vector field. 
We use the identity:
$$\Div \Vs = \Grad \Vs : \boldsymbol{I}.$$ 
Reynold's theorem for a scalar field $f$ defined in a volume $\mathscr{V}_t$ of a mixture with velocity $\bv$ reads
\begin{equation}
    \frac{d}{dt}\int_{\mathscr{V}_t} f dv  = \int_{\mathscr{V}_t} \partial_{t} f dv + \int_{\partial \mathscr{V}_t}  f \,\bv \cdot \bn da
    = \int_{\mathscr{V}_t} \bigg\{ \partial_{t} f + \ddiv (f \bv) \bigg\} dv \, .
    \label{eq:Appendix2}
\end{equation}
\\ The derivatives of the isochoric invariants with respect to $\bC$ are
\begin{subequations}
\begin{align}
    & \frac{\partial \bar{I}_{1}}{\partial \bF} = -\frac{2}{3}\bar{I}_{1}\FInvT + 2J^{-2/3}\bF,
    \label{Conto_01a} \\
    & \frac{\partial \bar{I}_{2}}{\partial \bF} = -\frac{4}{3}\bar{I}_{2}\FInvT + 2J^{-4/3}(I_{1}\bF - \bF \bC),
    \label{Conto_01b} \\
    &\frac{\partial J}{\partial \bF} = J \FInvT
    \label{Conto_01c}
\end{align}
\end{subequations}

\section*{Appendix B. Finite Element discretization}
\label{sec:appendixB}

The space discretization is achieved with finite elements, using linear interpolation elements for the pressure-like fields and quadratic interpolation elements for the displacement-like fields, in the form
\begin{align}
    & p (X,t) && = \bN^p (X) \, \bP(t) && = \sum_a N^p_a(X) \, p_a(t) \, , \nonumber \\
    & \eta (X) && = \bN^p(X) \, \bE && = \sum_a N^p_a(X) \, \eta_a(t) \, , \nonumber \\
    & \bu(X,t) && = \bN^u(X) \, \bU(t) && = \sum_a N^u_a(X) \, u_a(t) \, , \nonumber \\
    & \bxi(X) && = \bN^u(X) \, \bXi && = \sum_a N^u_a (X) \, \xi_a \, , \nonumber
\end{align}
where $\bP$, $\bU$, $\bE$, $\bXi$ are the arrays collecting the nodal pressures and displacements $p_a$, $u_a$ and the corresponding vector test $\eta_a$, $\xi_a$ components. The gradients become
\begin{align}
    & \Grad p(X,t) && =
    \nabla \bN^p(X) \, \bP(t) 
    && =
    \sum_a \nabla N_a^p(X) \, p_a(t) \, \nonumber \\
    & \Grad \eta(X) && =
    \nabla \bN^p(X) \, \bE 
    && = \sum_a \nabla N_a^p \, \eta_a \nonumber \\
    & \Grad \bu(X,t) && =
    \nabla \bN^u(X) \, \bU(t) && =
    \sum_a \nabla N_a^u \, u_a(t) \nonumber \\
    & \Grad \bxi(X) && =
    \nabla \bN^u(X) \, \bXi && =
    \sum_a \nabla N^u_a \xi_a \nonumber 
\end{align}
In formal notation, the space discretized equations read: 
\begin{equation}
    \left( \int_\Bref { \nabla \bN^p\, \bK \, \nabla \bN^p \, dV} \right) \bP(t)  +  
    \left( \int_\Bref { J \FInvT \bN^p   \nabla  \bN^u  } \, dV \right) \bU(t)
    =
    \int_{\partial_{\rm f} \Bref^{
    \rm q}} {\bN^p \, \bar \bQ(t) } \, dA \nonumber
\end{equation}
\begin{equation}
    - \left( \int_\Bref { J \FInvT \nabla \bN^u \bN^p  } \, dV \right) \bP(t)
    + 
    \int_\Bref {\nabla \bN^u  \Phis \Pel (\bU) } \, dV  = \int_{\partial_{\rm s} \Bref^{\rm t}} { \bN^u  \bar \bT(t) } \, dA  \,. \nonumber
\end{equation}
We introduce the global matrices and arrays as the assembling of the contribution of each element to the connected nodes
\begin{equation}
    \bM = \int_\Bref { \nabla \bN^p\, \bK \, \nabla \bN^p \, dV} \, ,
    \quad
    \bM^e_{ab} = \int_{\Bref^e} { \nabla N_a^p\, \bK^e \, \nabla N_b^p \, dV} \nonumber
    \label{permeabilityMatrix}
\end{equation}
\begin{equation}
    \bH = \int_\Bref { J \FInvT \bN^p   \nabla \bN^u  } \, dV \, ,
    \quad
    \bH^e_{ab} = \int_{\Bref^e} { J \FInvT N_a^p \, \nabla N_b^u \, dV} \nonumber
    \label{couplingMatrix}
\end{equation}
\begin{equation}
    \bB(t) = \int_{\partial_{\rm f} \Bref^{
    \rm q}} {\bN^p \, \bar \bQ(t) } \, dA \, ,
    \quad
    \bB^e_a(t) = \int_{\partial_{\rm f} \Bref^{
    \rm e\,q}} {N_a^p \, \bar \bQ^e(t) } \, dA  \nonumber
\end{equation}
\begin{equation}
    \bR^{\rm int}(\bU, t) = \int_\Bref {\nabla \bN^u  \Phis \Pel (\bU) } \, dV   \, ,
    \quad
    \bR^{\rm int \, e}_a(\bU, t) = \int_{\Bref^e} {\nabla N_a^u \Phis \Pel (\bU) } \, dV  
    \nonumber
\end{equation}
\begin{equation}
    \bR^{\rm ext}(t) = \int_{\partial_{\rm s} \Bref^{\rm t}} {\bN^u  \bar \bT(t) } \, dA   \, ,
    \quad
    \bR_a^{\rm ext \, e}(t) = \int_{\partial_{\rm s} \Bref^{\rm e \, t}} {N_a^u  \bar \bT^e(t) } \, dA   \, . \nonumber
\end{equation}

\section*{Appendix C. Augmented dissipation}
\label{sec:appendixC}

The rate of dissipation is augmented with the inclusion of a Lagrangian multiplier $p$
\begin{align}
    \Gamma (t) = & \left(\Ps -\Phis \Pel \right) : \dot{\bF}  + \Pf:\Grad\Vf + J \mf \cdot \left( \Vf -\Vs \right) \nonumber \\
    & + p \Div[(\phis \Vs+\phif \Vf)\, J \FInvT] \ge 0 \, .
\label{AugmentedDissipationInequality2}
\end{align}
The incompressibility constraint in $\Bref$ can be written using the Piola identity
\begin{equation}
    \Div[(\phis \Vs+\phif \Vf)\, J \FInvT] =
    \Div[\phis \Vs] : J \FInvT +
    \Div[\phif \Vf] : J \FInvT \, 
    \label{eq:manipulation1}
\end{equation}
which can be extended as
\begin{align}
    \Div[(\phis \Vs+\phif \Vf)\, J \FInvT] = & 
    \, J \, \Grad \phis \FInv \Vs +
    J \phis \Grad \Vs : \FInvT + \nonumber  \\
    & \, J \Grad \phif \FInv \Vf +
    J \phif \Grad \Vf : \FInvT  \,  
    \label{eq:manipulation2}
\end{align}
and, using $\phis +\phif = 1$ and $\Phis = J \phis$, $\phif = J \Phif$, we obtain
\begin{align}
    \Div[(\phis \Vs+\phif \Vf)\, J \FInvT] = & \, J \, 
    \Grad \frac{\Phis}{J} \FInv ( \Vs - \Vf) + \nonumber \\
    & 
    \, \Phis \FInvT : \Fdot + \Phif  \FInvT : \Grad \Vf \,.
    \label{eq:manipulation3}
\end{align}
Rearranging the terms in the dissipation rate
\begin{align}
    \Gamma (t) = & \left(\Ps -\Phis \Pel + p \Phis \FInvT\right) : \dot{\bF}  + \left( \Pf + p \Phif \FInvT \right) :\Grad\Vf  \, +  \nonumber \\
    &  J \left( \mf + p
    \Grad \frac{\Phis}{J} \FInv
    \right) \cdot \left( \Vf -\Vs \right) \ge 0 \, .
\label{AugmentedDissipationInequality3}
\end{align}
Under the hypotheses of \textit{hyperelastic solid phase} and \textit{macroscopically inviscid fluid phase}, the constitutive characterizations of $\Ps$ and $\Pf$ come out as in Equations \eqref{FirstPiolaKirchhoffSolid} and \eqref{FirstPiolaKirchhoffFluid}, respectively. The residual dissipation then reads
\begin{align}
    \Gamma_{\mathrm{res}} (t) =   J \left( \mf + p
    \Grad \frac{\Phis}{J} \FInv
    \right) \cdot \left( \Vf -\Vs \right) \ge 0 \, ,
\label{AugmentedDissipationInequality4}
\end{align}
and, by defining the \textit{dissipative part of the exchange of linear momentum between solid and fluid phase} as $\mfd:=\mf + p\Grad \frac{\Phis}{J} \FInv $, the residual dissipation in the reference placement can be written as  
\begin{align}
    \Gamma_{\mathrm{res}} (t) =   J \mfd \cdot \left( \Vf -\Vs \right) \ge 0 \, .
\label{AugmentedDissipationInequality5}
\end{align}
By rephrasing Equation \eqref{AugmentedDissipationInequality5} in terms of the material filtration velocity $\bQ = J\FInv\bq$, it yields
\begin{align}
    \Gamma_{\mathrm{res}} (t) =  \phif^{-1} \FT \mfd \cdot \bQ \ge 0 \, .
\label{AugmentedDissipationInequality6}
\end{align}
The simplest, non-trivial way to satisfy Equation \eqref{AugmentedDissipationInequality6} consists of providing a constitutive characterization of $\mfd$ so that the residual dissipation is a quadratic function of $\bQ$. Hence, we write $\mfd := \phif \FInvT\boldsymbol{K}^{-1}\bQ$, where $\boldsymbol{K}$ is a symmetric, positive definite tensor so that the residual dissipation is \textit{always satisfied} as $\Gamma_{\mathrm{res}} (t) =   (\boldsymbol{K}^{-1}\bQ) \cdot \bQ \ge 0$.

\bibliography{porousCornea}
\bibliographystyle{unsrt}

\end{document}